\documentclass{appolb}

\usepackage{graphicx}
\usepackage{amsmath}

\def\AA{A\hs-3.1mm$^{~^\circ}$}

\def\a{\alpha}

\def\g{\gamma}

\def\d{\delta}
\def\D{\Delta}
\def\e{\epsilon}

\def\h{\eta}

\def\m{\mu}

\def\O{\Theta}

\def\p{\pi}

\def\r{\rho}
\def\s{\sigma}
\def\S{\Sigma}

\def\vf{\varphi}

\def\f{\phi}
\def\x{\chi}

\def\w{\omega}
\def\W{\Omega}
\def\E{{\rm e}^+{\rm e}^-}

\def\ul{\underline}
\def\ol{\overline}
\def\hs{\hskip}
\def\vs{\vskip}

\def\\{\hfill\break}
\def\ni{\noindent}

\def\ran{\rangle}
\def\lan{\langle}

\def\lra{\longrightarrow}

\def\00{^o\hs-0.8truemm/\hs-0.8truemm_{oo}}
\def\12{^1\hs-0.8truemm/\hs-0.5truemm_2}
\def\32{\/^3\hs-0.9truemm/\hs-0.5truemm_2}

\newcommand{\Pom}{I$\!$P}
\newcommand{\mi}{\!-\!}

\def\ifmath#1{\relax\ifmmode #1\else $#1$\fi}%

\newcommand{\beq}{\begin{equation}}
\newcommand{\eeq}{\end{equation}  }
\newcommand{\beqa}{\begin{eqnarray}}
\newcommand{\eeqa}{\end{eqnarray}  }

\def\ifmath#1{\relax\ifmmode #1\else $#1$\fi}%

\def\rB{\ifmath{{\mathrm{B}}}}

\def\rd{\ifmath{{\mathrm{d}}}}

\def\rE{\ifmath{{\mathrm{E}}}}

\def\rK{\ifmath{{\mathrm{K}}}}

\def\rN{\ifmath{{\mathrm{N}}}}
\def\rp{\ifmath{{\mathrm{p}}}}

\def\rS{\ifmath{{\mathrm{S}}}}

\def\rT{\ifmath{{\mathrm{T}}}}

\def\lab{\ifmath{{\mathrm{lab}}}}

\def\tot{\ifmath{{\mathrm{tot}}}}

\def\lan{\langle}
\def\ran{\rangle}
 
\newcommand{\px}{\ensuremath{p_\mathrm{x}}}
\newcommand{\py}{\ensuremath{p_\mathrm{y}}}
\newcommand{\pz}{\ensuremath{p_\mathrm{z}}}
\newcommand{\mT}{\ensuremath{m_\mathrm{T}}}%
\newcommand{\pT}{\ensuremath{p_\mathrm{T}}}%
\newcommand{\taumodel}{$\tau$-model}

\newcommand{\chisq}{\ensuremath{\chi^2}}%

\begin{document}

\title{Hadron Correlations\\
at Energies from GeV to TeV%
}
\headtitle{Hadron Correlations}
\author{W. KITTEL
\address{Institute for Mathematics, Astrophysics and Particle Physics, Radboud University Nijmegen/Nikhef, Nijmegen, Netherlands 
\vskip 0.5 cm
{\sl Dedicated to Andrzej Bialas in honour of his 80th birthday}}}
\headauthor{W. Kittel} 
\maketitle
\begin{abstract}
One of the central issues in High Energy Physics is the close interchange between Theory and  Experiment. Ever since I know Andrzej Bia\l as, I kow him as one of the theorists most interested in   experimental data. This has naturally led to continuous
 fruitful contacts.\\
Even though we have been working somehow together since about 1968, we so far have only one single publication in common. 
This was back in 1969 and it was on means  to efficiently study what we then called (exclusive) Multihadron Final States. At that time this meant 3- or at best 4-particle final states of two-hadron  collisions at cms energies of some 4 GeV (not TeV!). The field of multiparticle dynamics was in fact the domain of Polish high-energy physicists. The first of a very successful (and still lasting) series of annual International Symposia on Multiparticle Dynamics was organized in Paris in 1970, but essentially by Polish 
physicists. Andrzej himself was not attending, but it was he who organized the third in these series in (of course) Zakopane.\\
Since heavy-ion collisions, another field of major interest for Andrzej, will be covered by others, I here will restrict 
myself mainly to the collisions of two elementary particles.
\end{abstract}


\newpage
\section{Exclusive longitudinal phase space analysis and its variables}

One of our common interests at that time was the so-called Longitudinal Phase Space (LPS) Analysis of 3- and 4- or even 5-particle final states and I will first try to recall the ideas behind that. 

The most complete way to study a so-called exclusive reaction of multiplicity $n$
\beq
A + B \lra C_1 + C_2 + \dots + C_n  
\eeq
is to look at the differential distribution of its matrix element in full phase space. This, however, requires a $3n\mi4$ dimensional analysis ($(3n\mi5)$-dimensional if the incident particles are unpolarized) and becomes increasingly impossible with increasing $n$. 

Nature helps: at low cms energies, the vast majority of collisions is "soft", i.e. leads to low transverse (with respect to the collision axis) momenta of final state particles, largely independent of the nature of the particle, the multiplicity $n$ and the cms energy $s^{\12}$. On the other hand, longitudinal (along the collision axis) momenta are unlimited (i.e. limited only by phase space) and depend strongly on the nature of the particle, the multiplicity and the energy.

In elastic and other two-particle production collisions, one is used to distinguish between forward and backward scattering. An extension of this classification to multiparticle final states is an analysis in just longitudinal phase space (LPS) \cite{1,2}. Then, each individual reaction of type (1) is represented by a point  with coordinates ($p_{\| 1}, \dots , p_{\| n}$) in a now only 
$n$-dimensional euclidean space $S_n$. Conservation of longitudinal momentum in the cms,                                                                          

\beq
\sum^n p^{*}_{\| i}=0\ ,
\eeq
defines LPS  as an $(n\mi1)$ dimensional hyperplane $L_{n-1}$. Furthermore, because of conservation of cms energy $s^{\12}$

\beq
\sum^n_{i=1}(m^2_i+ \rp^2_{\rT i}+p^{*2}_{\| i})^{\12}=s^{\12}\ .
\eeq
All points with equal transverse momentum  $|\rp_{\rT i}|$  lie on an $(n\mi2)$ dimensional hypersurface $K_{n-2}$   
defined  by  (3).  For the case of a transverse mass  $m_{\rT i }= (m_{i}^2+\rp_{\rT i}^2)^{\12} = 0$,   (3) reduces to
\beq
{\sum^n_{i=1}} |p^*_{\| i}| = s^{\12}
\eeq 
and defines a regular polyhedron $H_{n-2}$.  For $n=3$, this is the Van Hove Hexagon shown in Fig.~1 together with the 
one-dimensional manifold $K_1$. For $n=4$, the polyhedron $H_2$ is the cuboctrahedron celebrated in  Fig.~2.

A typical three-particle distribution in LPS for the final state of reaction
 \beq
\pi^-\rp \to \rp\pi^-\pi^0
\eeq                                                       
at an incident lab momentum of 16 GeV/$c$ is given in Fig.~3 \cite{3}. 

\begin{figure}[htb]
\begin{minipage}[t]{6cm}
\includegraphics[width=6cm]{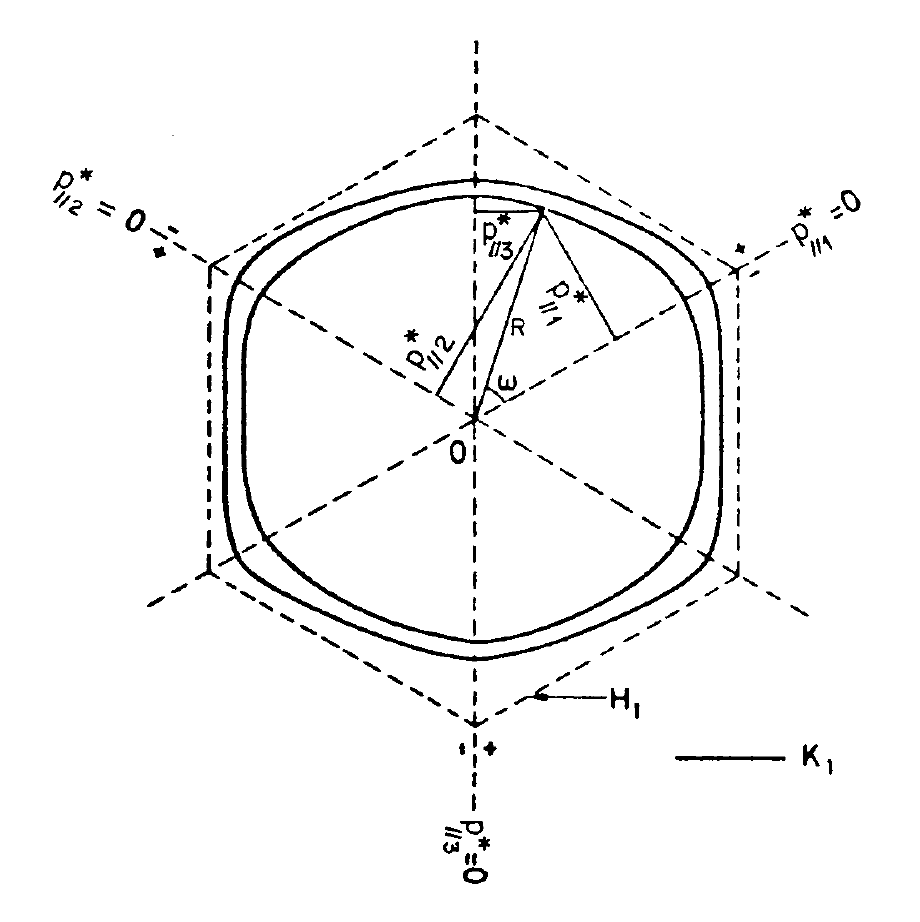}
\caption{Longitudinal phase space plot (Van Hove Hexagon) for the final state $\p\p$N\ at cms energy of 
$s^{\12}=4$ GeV. The innermost full line is $K_1$ for transverse 
momenta of 0.4, 0.4, and 0.5 GeV/$c$, respectively, 
while the outer one is $K_1$ for 
vanishing  transverse momenta. The dashed line represents the hexagon $H_1$ \cite{1}.
}

\end{minipage}
\hskip 5mm
\begin{minipage}[t]{6cm}
\includegraphics[width=6cm]{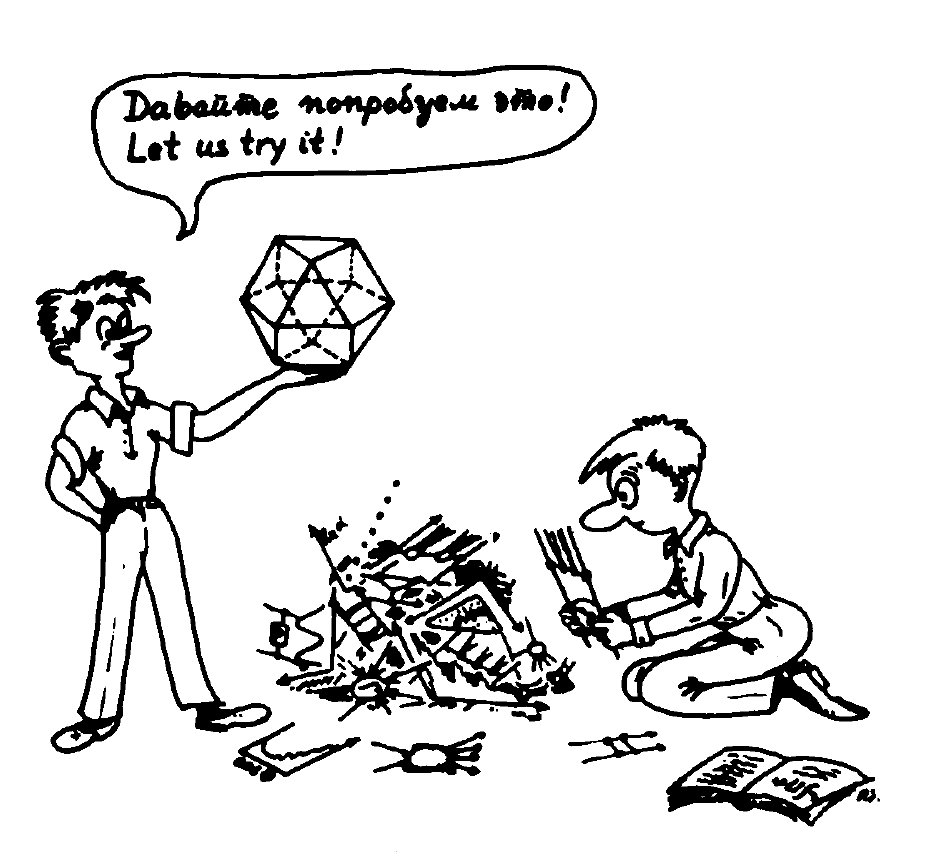}
\caption{The polyhedron $H_2$ for a four-particle final state (cartoon by R.Sosnowski).}
\end{minipage}
\end{figure}

\begin{figure}[htb]
\begin{minipage}[c]{6cm}
\includegraphics[width=6cm]{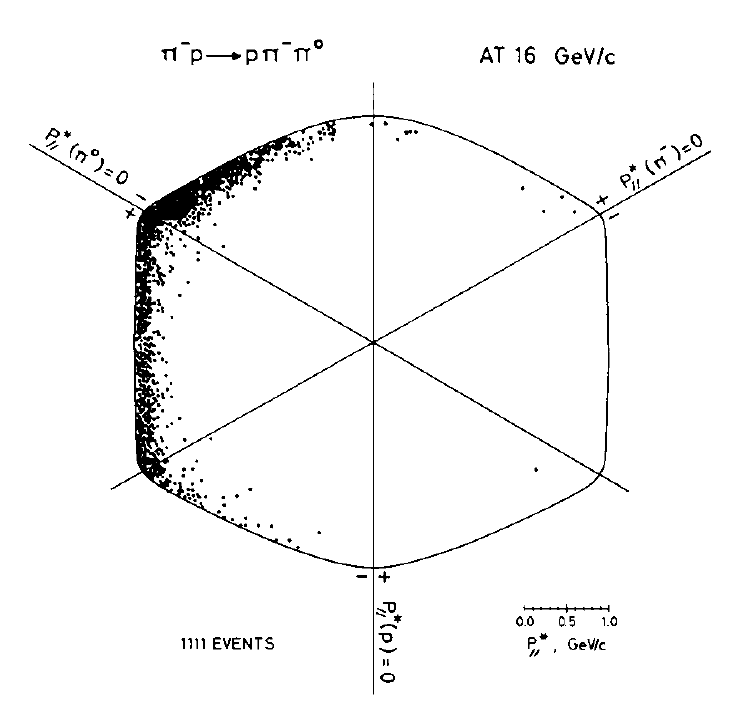}  
\caption{Distribution of final state points for the reaction $\p^-\to \p^-\p^0\rp$ at incident lab momentum 
of 16 GeV/$c$ \cite{3}.}
\end{minipage}
\hskip4mm
\begin{minipage}[c]{6cm}
\includegraphics[width=5cm]{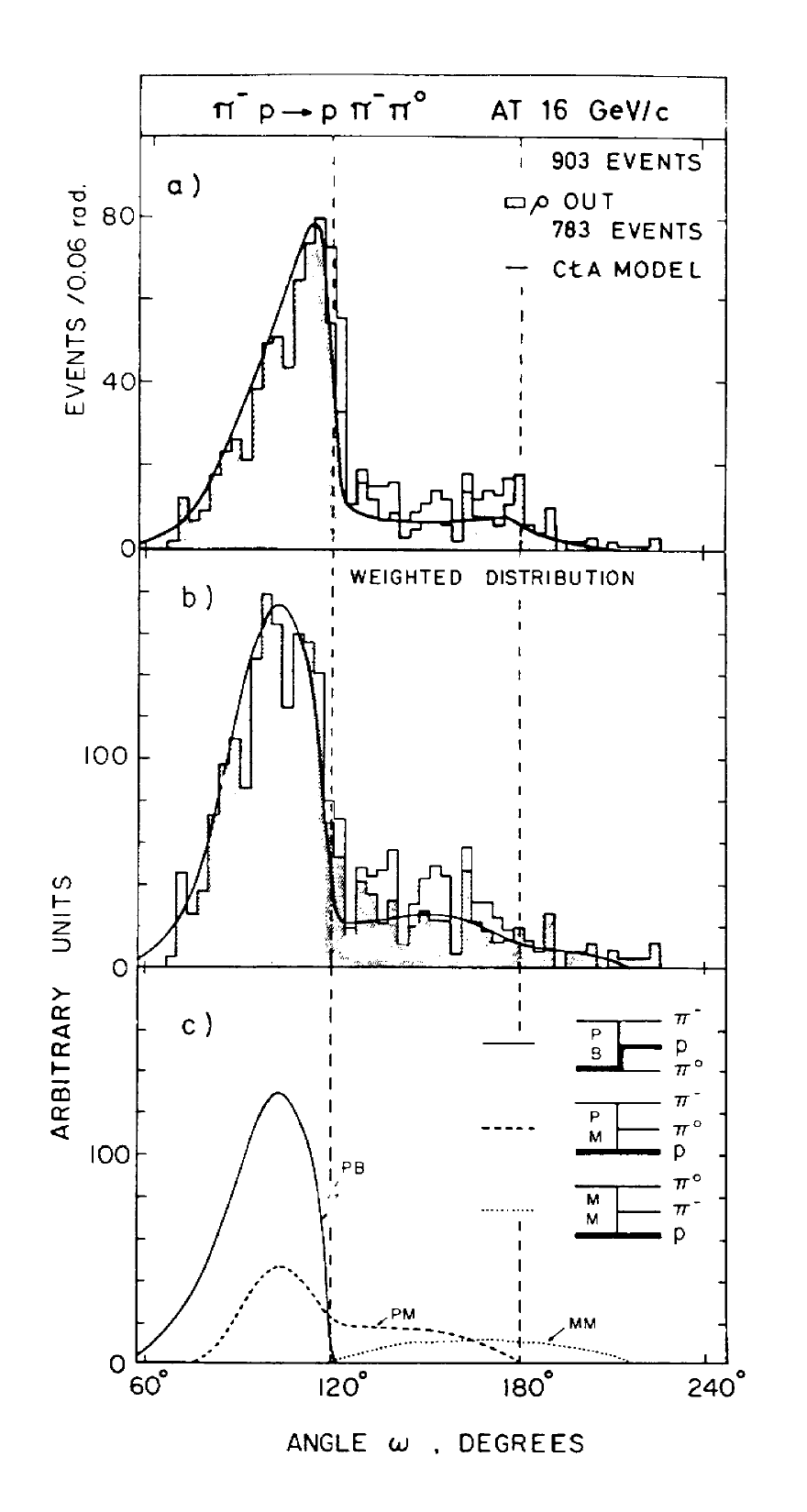}
\end{minipage}
\caption{(a) Distribution against the angle $\w$ of final state points for the reaction $\p^-\rp \to \p^-\p^0\rp$ at incident lab momentum of 16 GeV/$c$ \cite{3}. (b) same after correcting for non-constant phase space effects. 
The solid lines are the distributions according to the C\L A model \cite{4} normalized to the data after exclusion of the sharp 
$\r^-$-resonance. The individual C\L A exchange graphs considered and their contributions to the total distribution are given 
in sub-figure (c).}
\end{figure}

The distribution of (5) against the angle $\w$ is given in Fig.~4, before and after correcting for phase space effects (sub-figure (a) and (b), respectively). According to the definition in Fig.~1, 
the $\w$ region considered ($60^\circ<\w<120^\circ)$ corresponds\  to the hemisphere of LPS in which the proton is backward in the cms. The $\pi^0$\ is taken to be longitudinally at "rest" at $\w=120^\circ$, the $\pi^-$ at $\w=180^\circ$.                                                                                                                                                                                                                                                                        Peaks in the\ (model-independent!) experimental data (histograms) indicate strong correlations between particles in the final state, in particular in the region $60<\w<120^\circ$.

As a demonstration of how one can  use  LPS to test the success of theoretical models, the so-called C\L A model \cite{4}
of that time was used. It is a Reggeized form of a multiperipheral model, in which the amplitude is treated as an incoherent sum of contributions from various multiperipheral graphs. For the reaction studied in Fig.~4, they are given in sub-figure (c) together with their contributions. The full line in (a) and (b) corresponds to their incoherent sum.
After exclusion of a sharp $\r$-resonance, the model can describe the overall distribution of Fig.~4 surprisingly well. From the contribution of the graphs in sub-figure (c) we can see that vacuum exchange, commonly called \Pom(omeron) exchange, on the upper vertex essentially determines the shape of the distribution.

Turning back to a model independent data analysis, we investigate the energy dependence of the distribution and its shape in 
Fig.~5 according to its parametrization $\s(p_{\lab}) \propto p_{\lab}^{-N}$. As shown in sub-fibgure (b), $N$ is indeed close to zero for $60^\circ<\w<120^\circ$,\ in agreement with \Pom\ exchange (diffraction dissociation)  in that region \cite{5}.        
                                    
\begin{figure}[htb]
\centerline{%
\includegraphics[width=7cm]{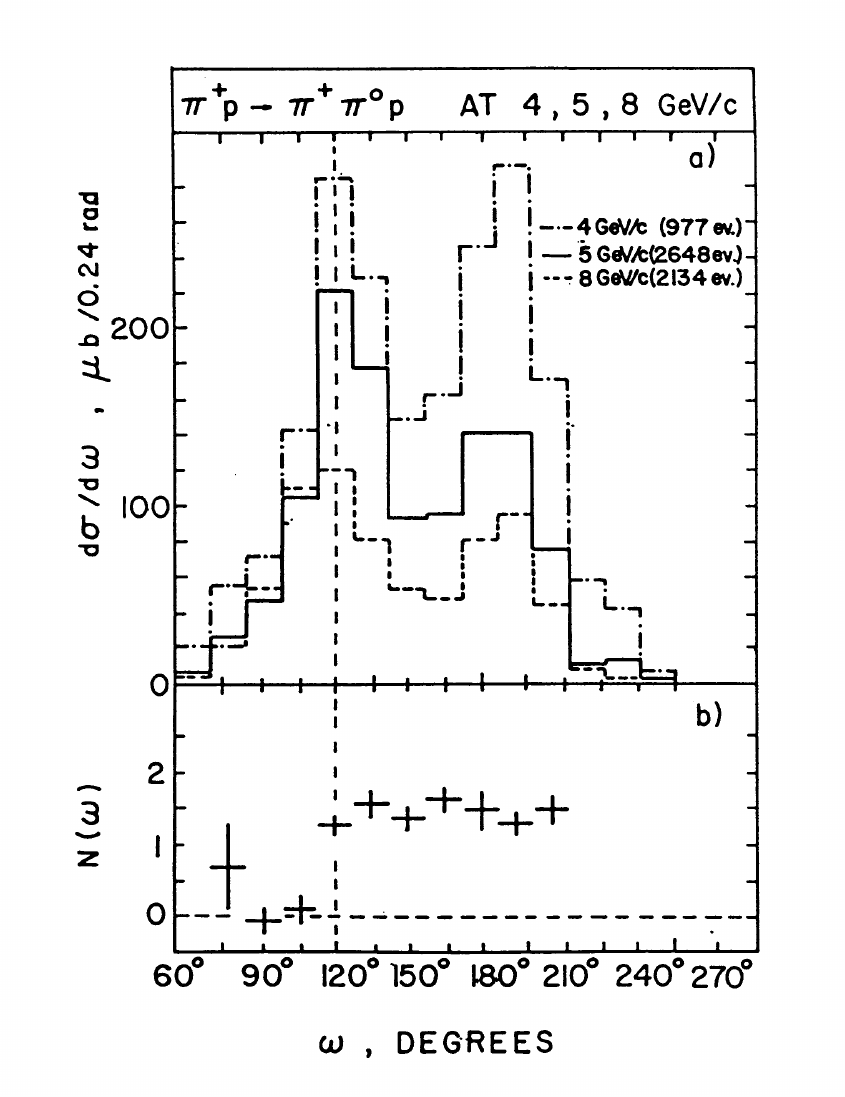}}
\caption{(a) $\w$-distribution for the reaction $\p^+\rp \to \p^+\p^0\rp$  at incident lab  momentum of 4, 5 
and 8 GeV/$c$ . (b) Exponent $N$ as a function of $\w$ for the same reaction \cite{5}.}
\end{figure}

Where, however, is the $\D$\ resonance? Unlike the incident proton, it has isospin $I=\32$ and can, therefore, not be produced via vacuum exchange.  

One way to look for it, is the so-called prism plot [6] ingeniously combining the advantages of the angle $\w$ along the z-axis with the subsystem masses given in a Fabri-Dalitz plot (triangle) at the basis (Figs.~6 and 7). The separation into individual mechanisms, each corresponding to a straight section of the tube within the prism, is better than in its projection onto the 
z-axis or the basis in the xy-plane. The mass of the ($\rp\p^+$)-subsystem is plotted in Fig.~8, for all events (sub-figure (a)) and for events in the corresponding section of the tube (sub-figure (b)). In the latter, the $\D^{++}$ is well separated from the background still present in (a).
    
\begin{figure}[htb]
\begin{minipage}[t]{6cm} 
\centerline{%
\includegraphics[width=6cm]{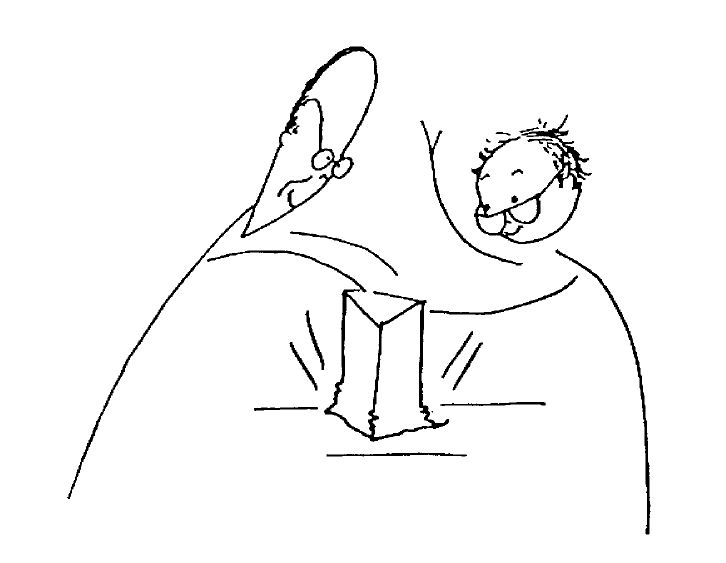}}
\caption{The prism plot as constructed by pulling a Fabri-Dalitz plot out in the direction of the Van Hove angle $\w$ (cartoon by Suzy Smile).}
\end{minipage}
\hskip4mm
\begin{minipage}[t]{6cm} 
\centerline{%
\includegraphics[width=6cm]{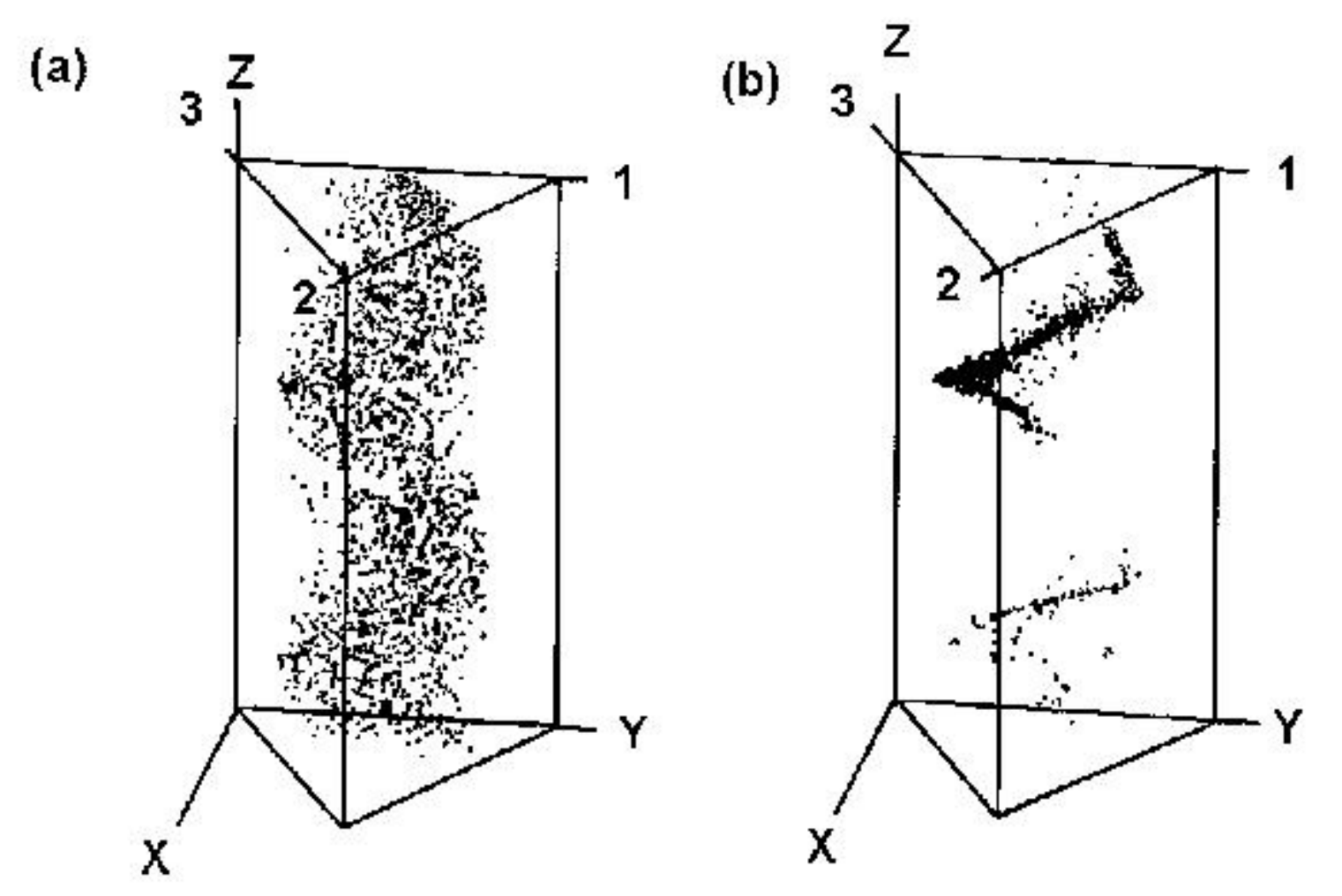}}
\caption{Prism plot for $\p^+\rp \to \p^+\p^0\rp$  at 3.9 GeV/$c$. (a) invariant phase space and (b) experimental
 data \cite{6}.}
\end{minipage}
\end{figure}

\begin{figure}[htb]
\includegraphics[width=6.5cm]{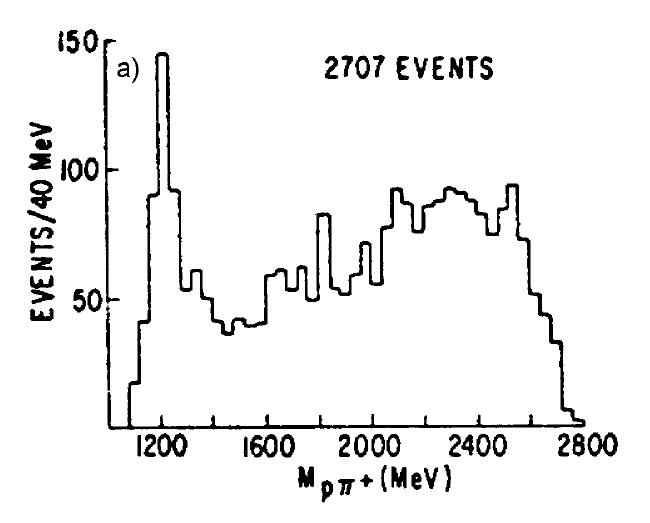} \hskip 1mm \includegraphics[width=6.5cm]{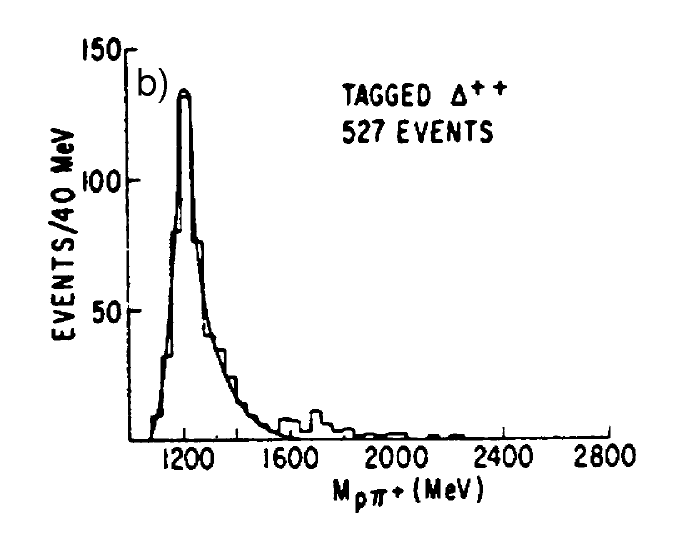} 
\caption{Effective mass of the $(\rp\p^+)$-subsystem for the reaction $\p^+\rp\to \p^+\p^0\rp$ at incident lab  
momentum of 3.9 GeV/$c$, for (a) all events and (b) for events in the corresponding section of the tube in the prism plot \cite{6}.}
\end{figure}

\begin{figure}[htb]
\centerline{%
\includegraphics[width=6cm]{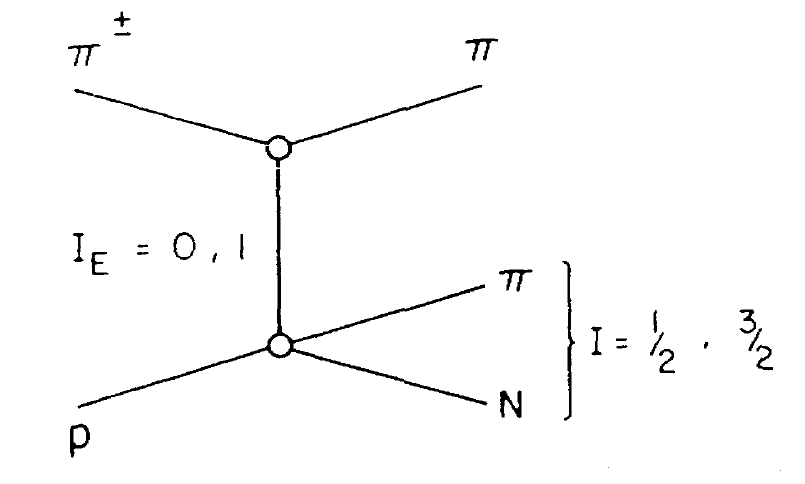}}
\caption{The three diagrams corresponding to the amplitudes specified by the exchanged isospin $I_\rE$ 
and the isospin $I$ of the $(\rN\p)$ system. (Note that the combination $I_\rE=0, I=\32$ is excluded from isospin conservation.)}
\end{figure}
    
\begin{figure}[htb]
\centerline{%
\includegraphics[width=10cm]{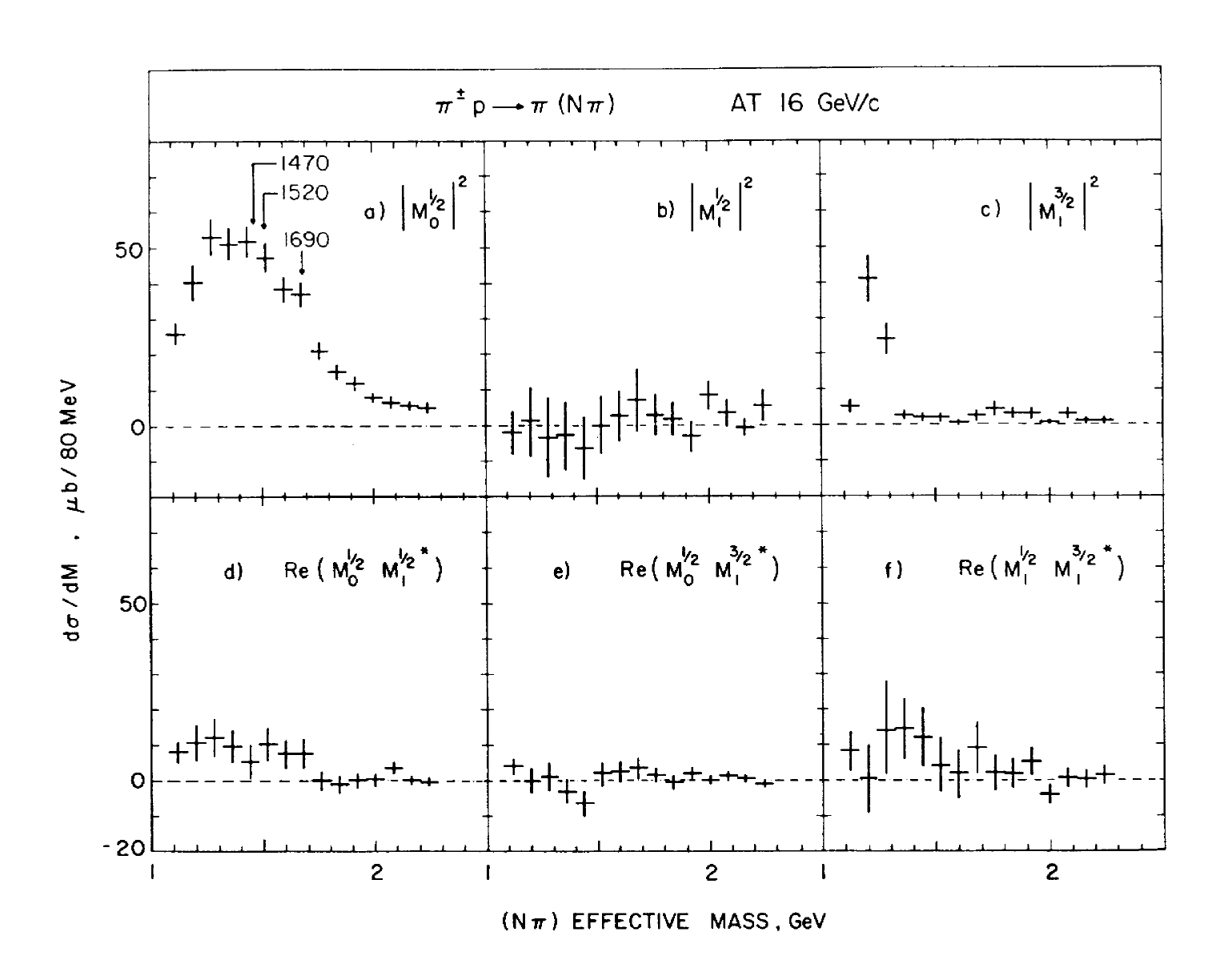}}
\caption{Squared amplitudes $|M_I^{I_\rE}|^2$ and their interference terms as functions of the $(\rN\p)$ mass obtained from the reactions $\p^\pm \rp \to \p\p\rN$ at 16 GeV/$c$ \cite{7}.}
\end{figure}
    
Another way of extracting a pure $\D$ signal is a separation of the 
 isospin matrix element according to the graph in 
Fig.~9. Since isospin exchange $I_\rE=0$ is excluded for the production of the  $I=\32\  (\rN\p)$-system, we are left with three matrix elements and their interferences. Their squares, respectively real parts, can readily be extracted model independently 
from combinations of the 6 measurable (of the 7 possible) final states of $\p^\pm\rp$ reactions. They are given in Fig.~10 as a function of the $(\rN\p)$ effective mass. While a wide diffractive shoulder is observed in sub-figure (a), a sharp and well 
separated $\D$ can be seen in subfigure (c).

Overlap between different sub-systems is a problem, in particular in the determination of spin-parity of a particular sub-system (partial wave analysis). However, interference also provides a unique possibility to study the relative phase between overlapping amplitudes. In such  a  study,  all  mechanisms  contributing  to  a  particular  few-body  final  state  have  to  be treated simultaneously in an iterative and interactive computer analysis. A beautiful method allowing that is the so-called Analytical Multichannel Analysis \cite{8}.

The method has successfully been applied to 30 000 events of the final state of $\rK^-\rp\to \ol\rK^0\p^-\rp$
at 4,2 GeV/$c$ \cite{9}. As the four variables needed to describe a three-particle final state, the effective mass 
$M$ has been used for the sub-system considered, the invariant four-momentum $t'$ in its production and the two angular variables $\O$ and $\f$ of its decay.

\begin{figure}[t]
\centerline{%
\includegraphics[width=11cm]{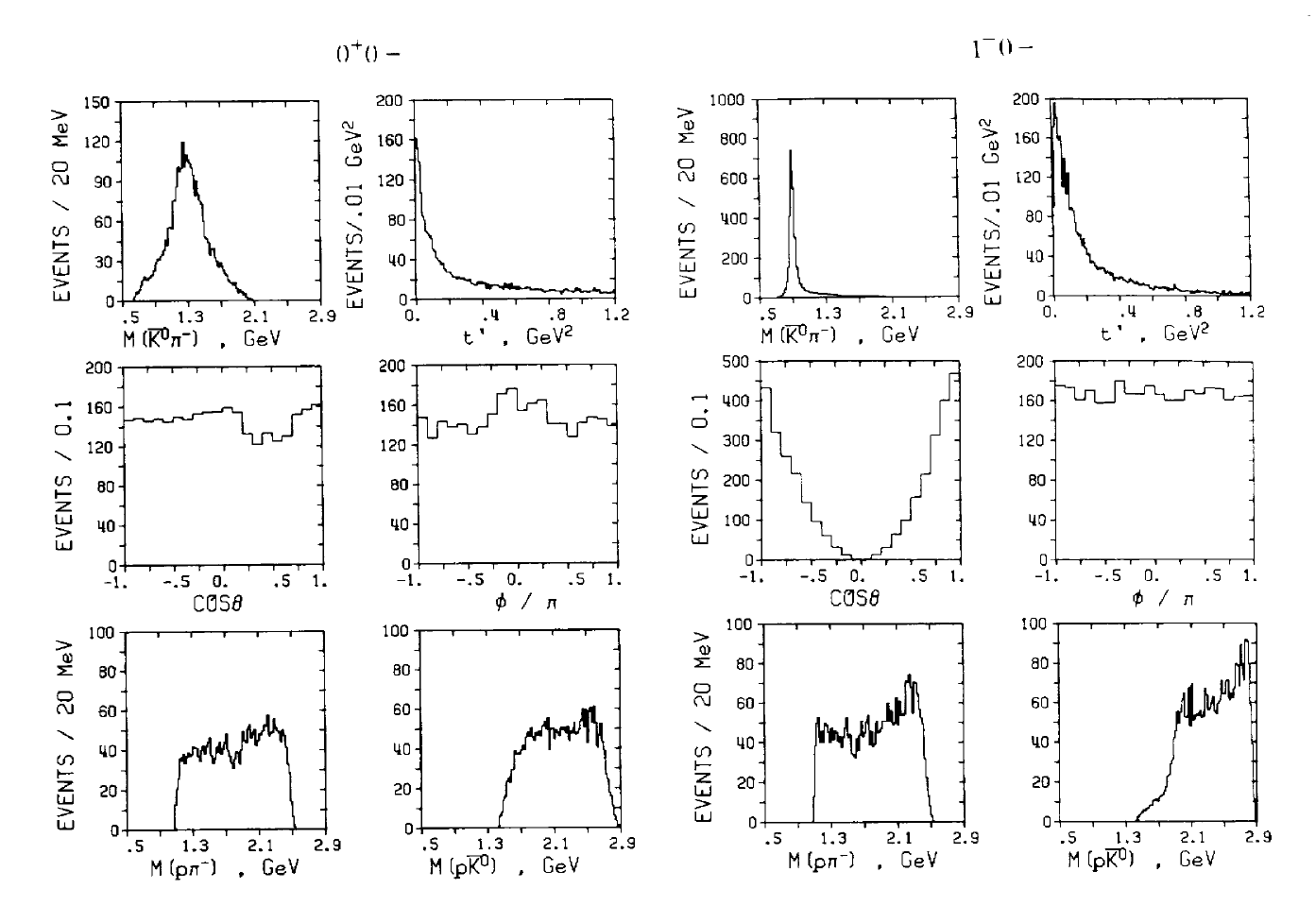}}
\caption{Effective mass distribution of the $(\ol\rK^0\p^-)$ system, four-momentum transfer $t'$ from 
initial to final state proton, decay angles of the $(\ol\rK^0\p^-)$ system, and effective mass of the $(\rp\p^-)$- and
 $(\rp \ol\rK^0)$-systems for the 
$0^+0-$ and $1^-0-$ $(\ol\rK^0\p^-)$  samples after iteration 9 \cite{9}.}
\end{figure}
                      
As examples for the results after 9 iterations, Figs. 11 and 12 correspond to the $(\ol\rK^0\p^-)$ S-wave and its P-waves 
 $1^-0-$, $1^-1-$ and $1^-1+$. Except for the S-wave which is not yet flat, the angular distributions correspond to the particular wave and are as expected. Of particular interest are the differences in the four-momentum $t'$ distributions for the three P-waves, typical for pseudo-scalar and vector exchange, respectively.

Striking is the difference in the reflection into the $(\rp\pi^-)$ and $(\rp\ol\rK^0)$ systems in the lowest row of Fig.~12. 
The Monte Carlo curve superimposed on the $(\rp\pi^-)$ mass distribution shows a two-peak reflection  from the $1^-1-$ 
wave. Just mind the enormous error introduced by the simple smooth hand-drawn background as used in earlier conventional analysis!

Very similar results are obtained \cite{9} for the three D-and even F-waves and for other sub-channels down to the $\00$ level of their contribution to the total final state, not detectable in earlier analysis.

\begin{figure}[t]
\centerline{
\includegraphics[width=11cm]{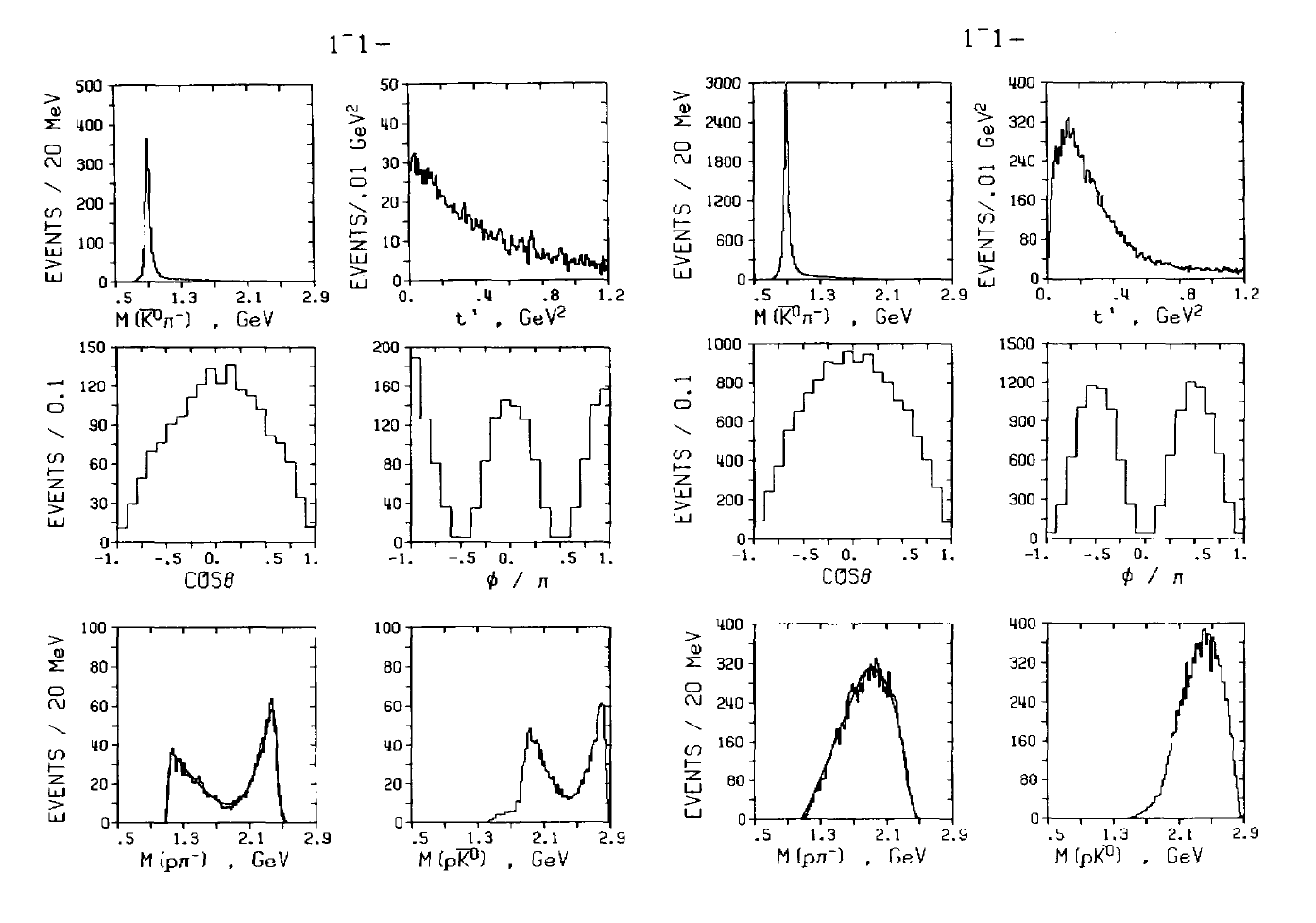}}
\caption{Same as Fig. 11, but for the $1^-1-$ and $1^-1+$ samples \cite{9}.}
\end{figure}
   
In conclusion from this section: With the help of model independent data analysis we have moved from analysis in Longitudinal Phase Space to a complete Multichannel Analysis. While the  LPS analysis has demonstrated strong correlation of final state particles, the increased number of variables of the prism plot could show overlap of these mechanisms in full phase space (not just in projections of it). These mechanisms can be separated by means of quantum numbers  as isospin, spin and angular momentum, and their interferences can be studied when all channels contributing to a final state are treated simultaneously. This analysis is particularly useful for the isolation of channels at the permille level of cross section or branching ratio.

What had we learned for the future? Experiments have to be {\it complete} in the sense that acceptance losses should be minimal and the four-vectors of all particles should be known. Furthermore, the analysis has to be done {\it iteratively and interactively}, i.e has to be guided by computer graphics.   
                                                
\section{Momentum correlations and density fluctuations}
\subsection{The formalism}
We start by defining symmetrized inclusive $q$-particle distributions
\beq
\r_q (p_1,\dots,p_q)= \frac{1}{\s_\tot} \frac{\rd \s_q(p_1,\dots,p_q)}{
\prod\limits^q_1 \rd p_q}\ \ ,
\eeq
\vs-3mm
\ni
where $\s_q(p_1,\dots,p_q)$ is the inclusive cross section for $q$
particles to be at $p_1,\dots,p_q$, irrespective of the presence and location
of any further particles, $p_i$ is the (four-) momentum of particle $i$ and
$\s_\tot$ is the total hadronic cross section of the collision under
study.
For the case of identical particles, integration over an interval
$\W$ in $p$-space  yields
\begin{eqnarray}
 &~&\int_\W \r_1(p) \rd p = \lan n\ran\ , \ \ \ \ \ 
 \int_\W \int_\W \r_2(p_1,p_2)\rd p_1\rd p_2 =
\lan n(n-1)\ran \ ,\nonumber \\
 &~&\int_\W \rd p_1 \dots \int_\W \rd p_q \r_q (p_1,\dots,p_q) =
\lan n(n-1)\dots (n-q+1)\ran \ ,
\end{eqnarray}
where $n$ is the multiplicity of identical particles within $\W$ in a 
given event and the angular brackets imply the average over the event ensemble.

Besides the interparticle {\it correlations} we are looking for,
the inclusive $q$-particle number densities $\rho_q(p_1,\dots,p_q)$ in general
contain ``trivial'' contributions from lower-order densities. 
It is, therefore, advantageous to consider a new sequence of functions
$C_q(p_1,\dots,p_q)$ as those statistical quantities which vanish whenever one
of their arguments becomes statistically independent of the others
\cite{kahn:uhlenbeck,huang,Mue71}:
\vs-3mm
\begin{eqnarray}
C_2(1,2)&=&\rho_2(1,2) -\rho_1(1)\rho_1(2)\ ,\\
C_3(1,2,3)&=&\rho_3(1,2,3)
-\sum_{(3)}\rho_1(1)\rho_2(2,3)+2\rho_1(1)\rho_1(2)\rho_1(3)\ ,
\label{a:4b}
\end{eqnarray}
\vs-3mm
\ni
etc. In the above relations, we have abbreviated $C_q(p_1,\dots,p_q)$ to\break 
$C_q(1,2,\dots,q)$; the summations indicate that all possible permutations 
must be taken. Expressions for higher orders can be derived from the related 
formulae given in~\cite{kendall}. 
Deviations of these functions from zero shall be addressed as {\it genuine}
correlations. 

It is often convenient to divide  the functions
$\rho_q$ and $C_q$ by the product of $q$ one-particle densities, which leads to
the  definition of the  normalized inclusive densities and correlations:
\begin{eqnarray}
R_q(p_1,\dots,p_q) &=& \rho_q(p_q,\dots,p_q)/\rho_1(p_1)\ldots
\rho_1(p_q),\label{3.8}\\
K_q(p_1,\ldots,p_q)& =& C_q(p_1,\ldots,p_q)/\rho_1(p_1)\ldots
\rho_1(p_q).\label{3.9}
\end{eqnarray}
In terms of these functions, correlations have been studied extensively for 
$q=2$. Results also exist for $q=3$, but usually the statistics (i.e. number 
of events available for analysis) are too small to isolate genuine 
correlations. To be able to do that for $q\geq 3$, one must apply factorial
moments $F_q$
defined via the integrals Eq.~(7), but in limited phase-space cells 
\cite{bialas,wolf96}.

\subsection{Density spikes}

To see whether it is worth the effort, we first look for
density fluctuations in single events, signalling high-order correlations.
A notorious JACEE event~\cite{Burn83} (Fig.~13a) at a pseudo-rapidity
resolution (binning) of $\d\h=0.1$ has
local fluctuations up to $\rd n/\rd\h\approx300$ 
with a signal-to-background ratio of about 1:1. An NA22 event
\cite{AdamPL185-87} (Fig.~13b) contains a ``spike" at a rapidity 
resolution $\d y=0.1$ of $\rd n/\rd y=100$, as much
as 60 times the average density in this experiment. 

\begin{figure}[htb]
\begin{minipage}[t]{6cm} 
\includegraphics[height=6.9cm]{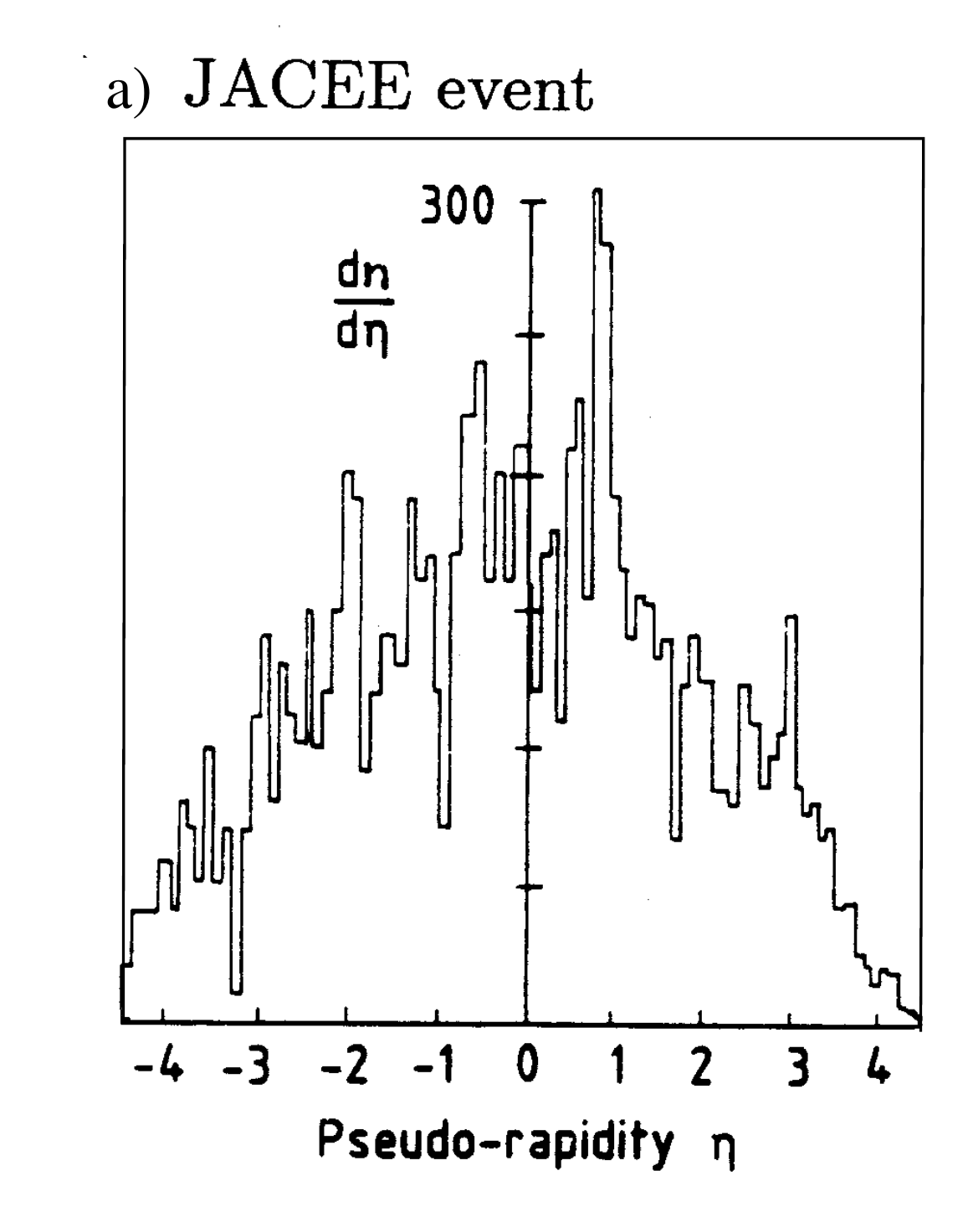}
\end{minipage}
\hskip -4mm 
\begin{minipage}[t]{6cm}
\includegraphics[height=6.9cm]{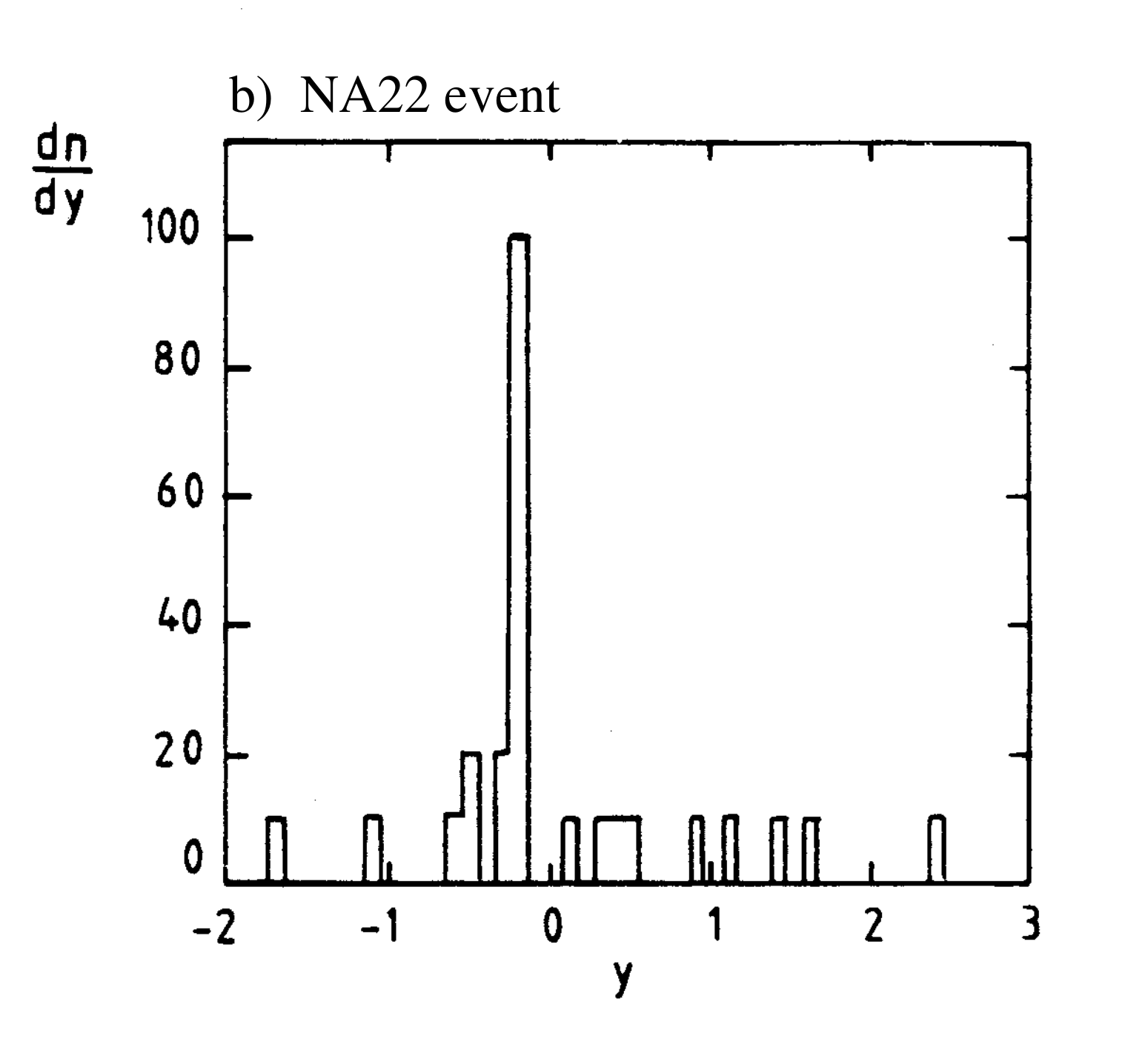}
\end{minipage}
\caption{a) The JACEE event \cite{Burn83}; b) The NA22 event \cite{AdamPL185-87}.}
\end{figure}

{Bia\l as} and Peschanski \cite{bialas} suggested that this type of spikes 
could be a 
manifestation of ``intermittency'', a phenomenon well known in fluid dynamics
 \cite{Zeld90}. The authors argued that if 
intermittency indeed occurs in particle production, large density 
fluctuations are not only expected, but should also
exhibit self-similarity with respect to the size of the
phase-space volume. 

In multiparticle experiments, the number of hadrons produced in a single 
collision is small and subject to considerable noise. To 
exploit the techniques employed in complex-system 
theory, a method had to be devised to separate fluctuations of purely 
statistical (Poisson) origin, due to finite particle numbers, from 
possibly self-similar dynamical fluctuations of the underlying 
particle densities. 
A solution, already used in quantum optics 
\cite{Bedard:67:1} and suggested for 
multiparticle production in~\cite{bialas}, consists in measuring 
$F_q(\d y)$ in given 
phase-space volumes (resolution) $\d y$ of ever decreasing size. 

\subsection{Power-law scaling}

Besides the property 
of noise-suppression, high-order factorial moments act as a filter 
and resolve the large-multiplicity tail of the multiplicity distribution.
They are thus  particularly sensitive to large density fluctuations 
at the various scales $\d y$ used in the analysis.
As shown in~\cite{bialas}, a smooth density distribution, which does 
not show any fluctuations except for the statistical ones, has the property 
of normalized factorial moments $F_q(\d y)$ being independent of the 
resolution $\d y$ in the limit $\d y\to  0$. On the other hand, if 
self-similar dynamical fluctuations exist, the $F_q$ obey the power law
\beq
F_q(\d y) \propto (\d y)^{-\f_q}\ , \ \ (\d y\to 0).
\label{12}
\eeq

Equation (\ref{12}) is a scaling law
since the ratio of the factorial moments at resolutions $L$ and $\ell$
\beq
R = \frac{F_q(\ell)}{F_q(L)} = \left(\frac{L}{\ell}\right)^{\f_q}
\eeq
only depends on  the ratio $L/\ell$, but not on $L$ and $\ell$,
themselves.

\begin{figure}[h]
\centerline{%
\includegraphics[width=7cm]{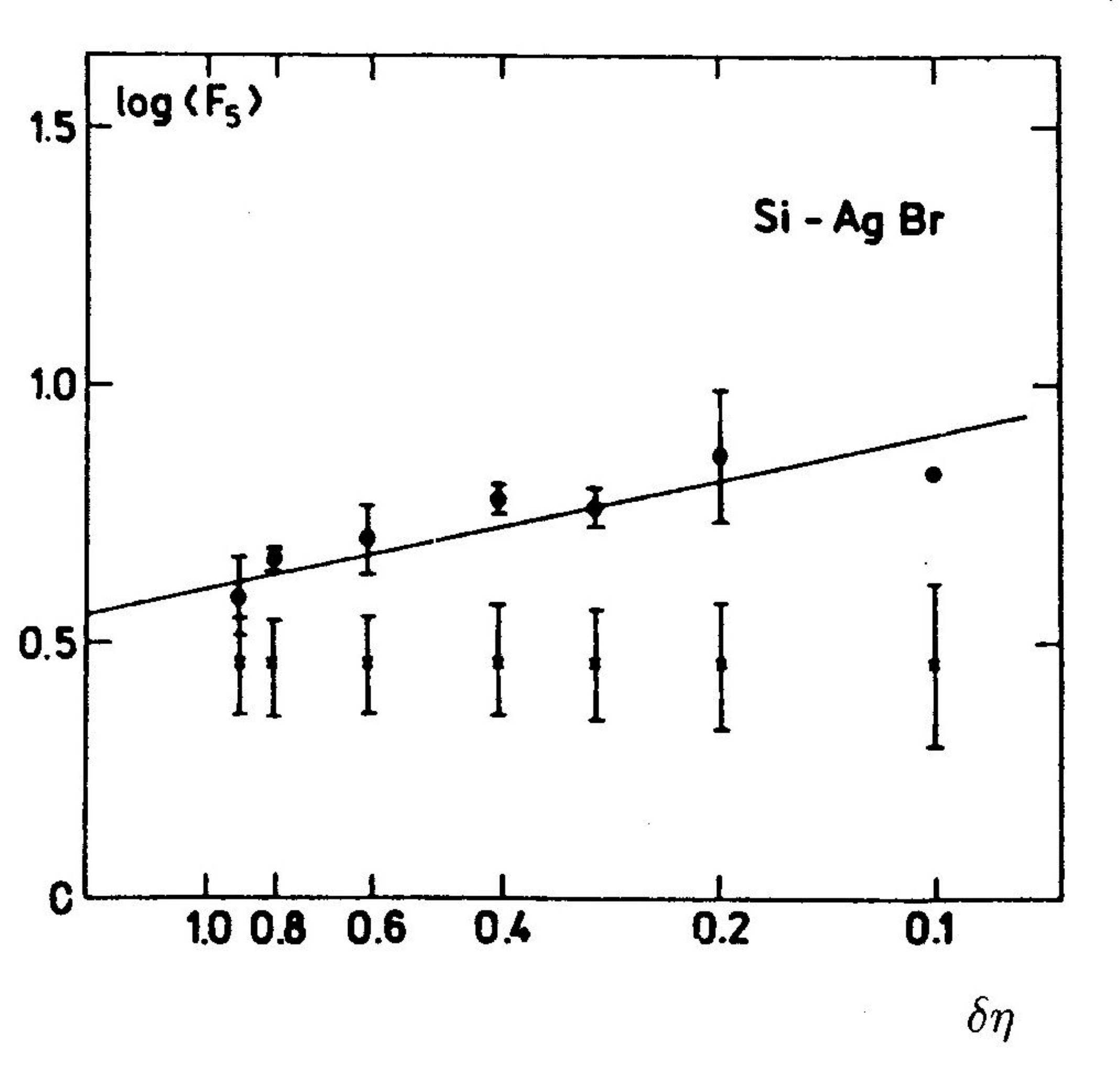}}
\caption{ log$F_5$ as a function of $-\log\d\h$ for the JACEE
event~\cite{Burn83} (full circles) compared to independent 
emission (small crosses) \cite{bialas}.} 
\end{figure}

In Fig.~14, log$F_5$ is plotted~\cite{bialas} as a function of -log 
$\d\h$ ($\h$ is the pseudorapidity) for the JACEE 
event. It is compared with an independent-emission
Monte-Carlo
model tuned to reproduce the average $\h$ distribution of 
Fig.~13 a) and the global multiplicity
distribution, but has no short-range correlations.
While the Monte-Carlo model indeed predicts constant $F_5$, the 
JACEE event shows a first indication for a linear increase, i.e. a possible 
sign of intermittency.

This observation was the trigger for a tremendous outburst of
 experimental research on all types of collisions from e$^+$e$^-$ to heavy
nuclei, all showing (approximate) power law scaling. An 118 page
summary including more than 300 references is given as chapters
7 and 10 in \cite{kitwolf2005}.
          
The powers $\f_q$ (slopes in a double-log plot) are related~\cite{LiBu89} 
to the anomalous (or co-) dimensions $d_q=\f_q/(q-1)$, a measure for the 
deviation from an integer dimension. ​     

Anomalous dimensions $d_q$ fitted over the (one-dimensional) 
range $0.1<\d y<1.0$ are compiled in Fig.~15~\cite{Bia1991}. They typically 
range from $d_q=0.01$ to $0.1$, which means that the 
fractal (R\'enyi) dimensions $D_q=1-d_q$ are close to one. 
The $d_q$ are larger and grow faster with increasing order $q$ in $\m$p and 
$\E$ (Fig.~15a) than in hh collisions (Fig.~15b) and are small and almost 
independent of $q$ in heavy-ion collisions (Fig.~15c). For hh collisions, 
the $q$-dependence is considerably stronger  for NA22 ($\sqrt{s}=22$ GeV, 
all $p_\rT$) than for UA1 ($\sqrt{s}=630$ GeV, $p_\rT>0.15$ GeV/$c$).

\begin{figure}[htb]
\centerline{
\includegraphics[width=12.5cm]{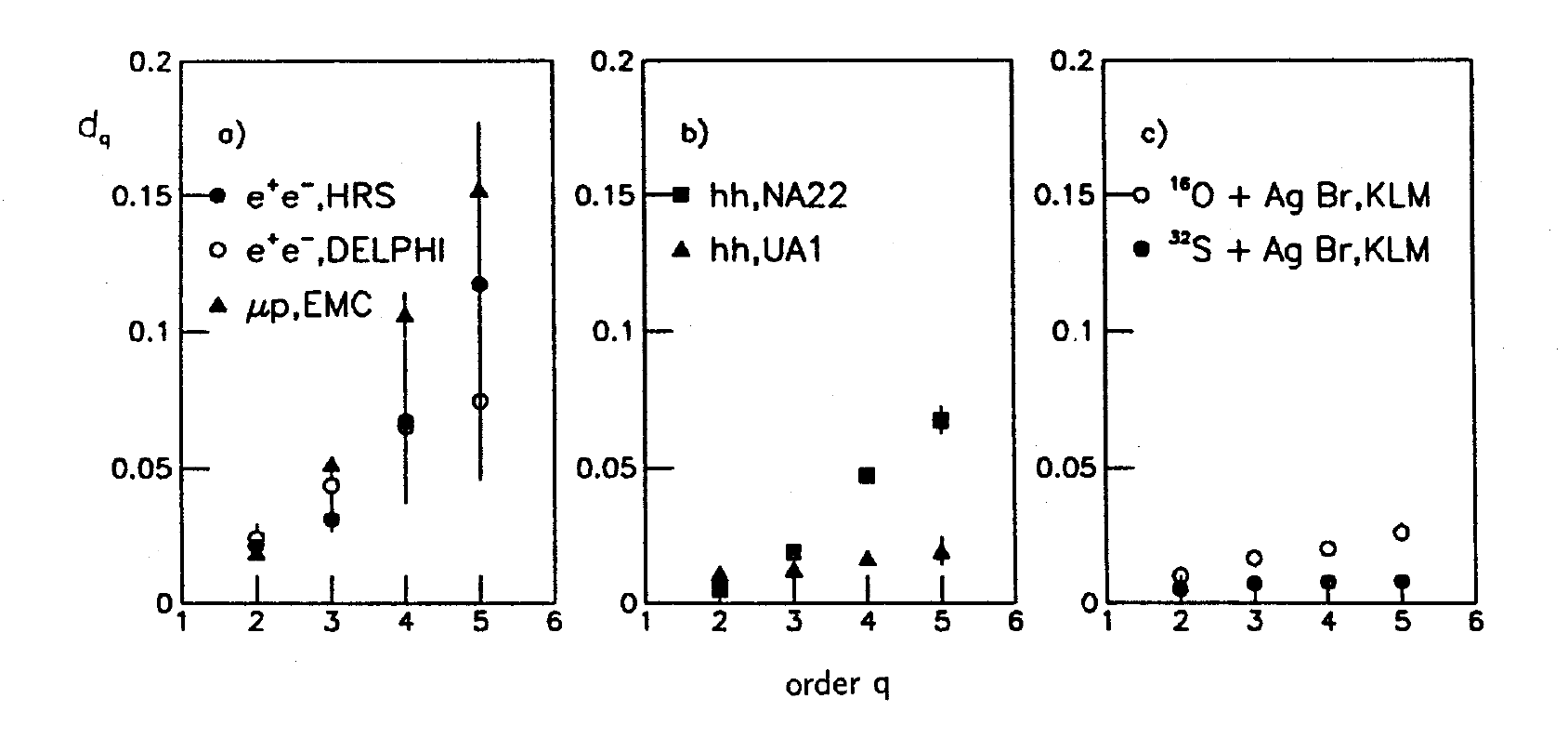}}
\caption{
Anomalous dimension $d_q$ as a 
function of the order $q$, for a) $\m$p and $\E$ collisions, b) NA22 and UA1, 
c) KLM~\cite{Bia1991}.}
\end{figure}

\subsection{Factorial Cumulants}

One further has to  stress the advantages of normalized factorial 
cumulants $K_q$ 
compared to factorial moments, since the former measure {\it genuine}
correlation patterns.

As an example, high statistics data of the OPAL experiment \cite{Sar00}
are given in Fig.~16 in terms of $K_q$, as a function of the number 
$M\propto 1/\d y$ of phase
space partitions for $q=3$ to 5. In the leftmost column, the one-dimensional
rapidity variable $y$ is used for the analysis. The data (black dots) show an
increase of $K_q$ with increasing $M$ for small $M$, but a saturation at
larger $M$. Even though weaker, some saturation still persists when the
analysis is done in the two-dimensional plane of rapidity $y$ and azimuthal
angle $\varphi$ (middle column), but approximate power-law scaling is indeed 
observed for the analysis in three-dimensional momentum space (right column). 
Thus, in high-energy collisions, fractal behavior is fully 
developed in three dimensions, while projection
effects lead to saturation in lower dimension.

\begin{figure}[htb]
\centerline{
\includegraphics[width=9cm]{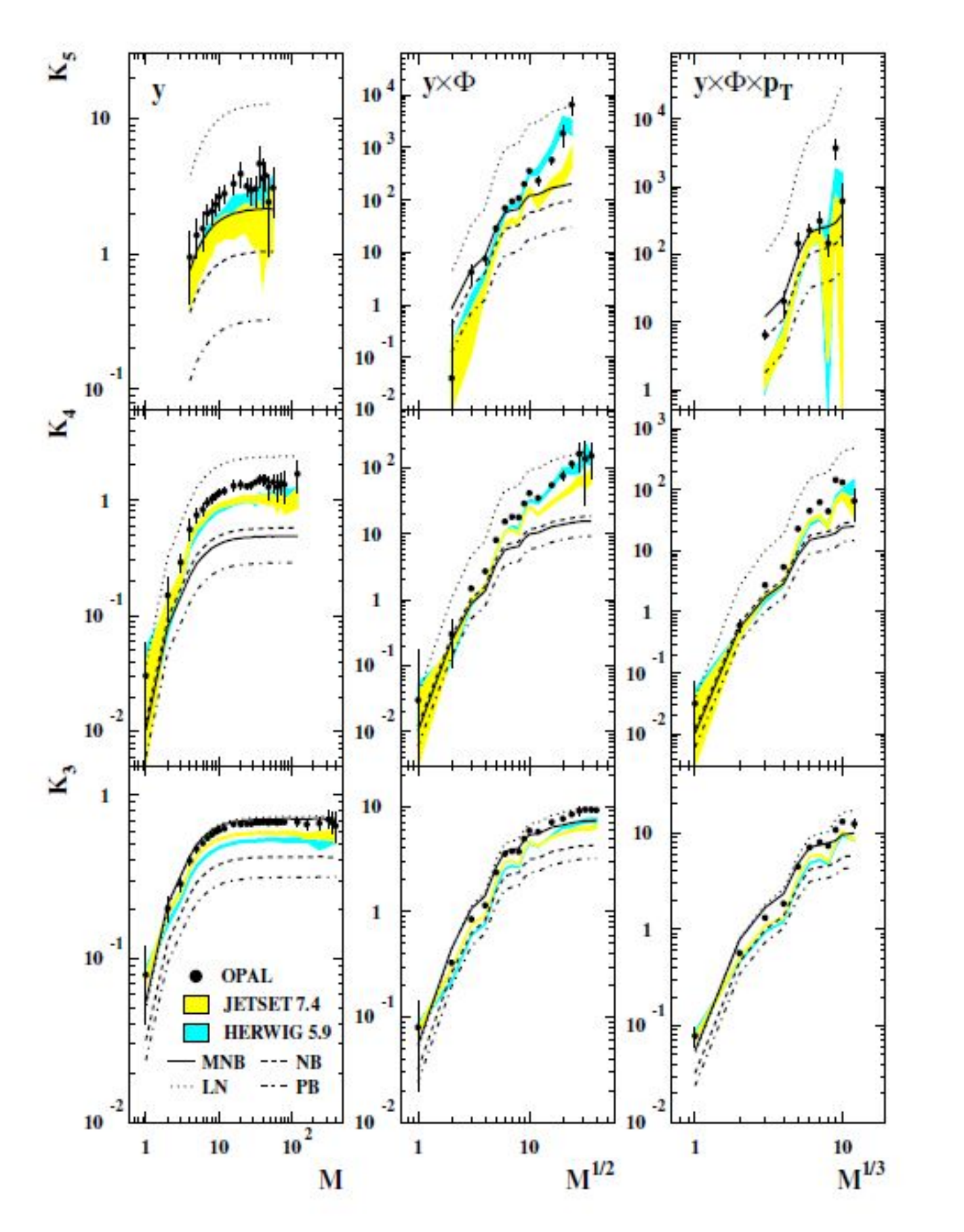}}
\caption{
Cumulants of order $q=3$ to 5 as a function of $M^{1/D}$ 
in comparison with the predictions of various multiplicity parametrizations 
and two Monte Carlo models \cite{Sar00}. }
\end{figure}

In Fig.~16, the data are also compared to a number of parametrizations 
of the multiplicity distributions, as well as to the Monte Carlo models
JETSET  and HERWIG. One can see that the fluctuations given by the 
negative binomial (NB) (dashed line) are weaker than observed in the data. 
Contrary to the NB, the log-normal (LN) distribution (dotted line) 
overestimates the cumulants, while these expected for a pure birth (PB)
process (dash-dotted) underestimate the data even more significantly than 
the NB. Among the distributions shown, a modified NB (MNB) gives the best 
results, even though significant underestimation is observed
also there. The Monte Carlo models do surprisingly well.

\subsection{Transverse-momentum dependence}
An interesting question is whether semi-hard effects
\cite{OchWo88-89}, observed to play a role in the transverse-momentum behavior
even at NA22 energies~\cite{Ajin87-2}, or low-$p_\rT$
effects ~\cite{VHov89,BialPL89} are 
at the origin of intermittency. A first indication for 
the latter comes from the most prominent NA22 spike event 
(Fig. 13b), where 5 out of 10 tracks in the spike have 
$p_\rT<0.15$ GeV/$c$.

In Fig.~17, NA22 data \cite{Ajin89-90} on $\ln F_q$ versus $-\ln\d y$ 
are given for 
particles with transverse momentum $p_\rT$ below and above 0.15 GeV/$c$, and 
with $p_\rT$ below and above 0.3 GeV/$c$. For particles with $p_\rT$ below the 
cut (left), the $F_q$ exhibit a far stronger $\delta y$ dependence than for 
particles with $p_\rT$ above the cut (right).

\begin{figure}[b]
\centerline{
\includegraphics[width=7cm]{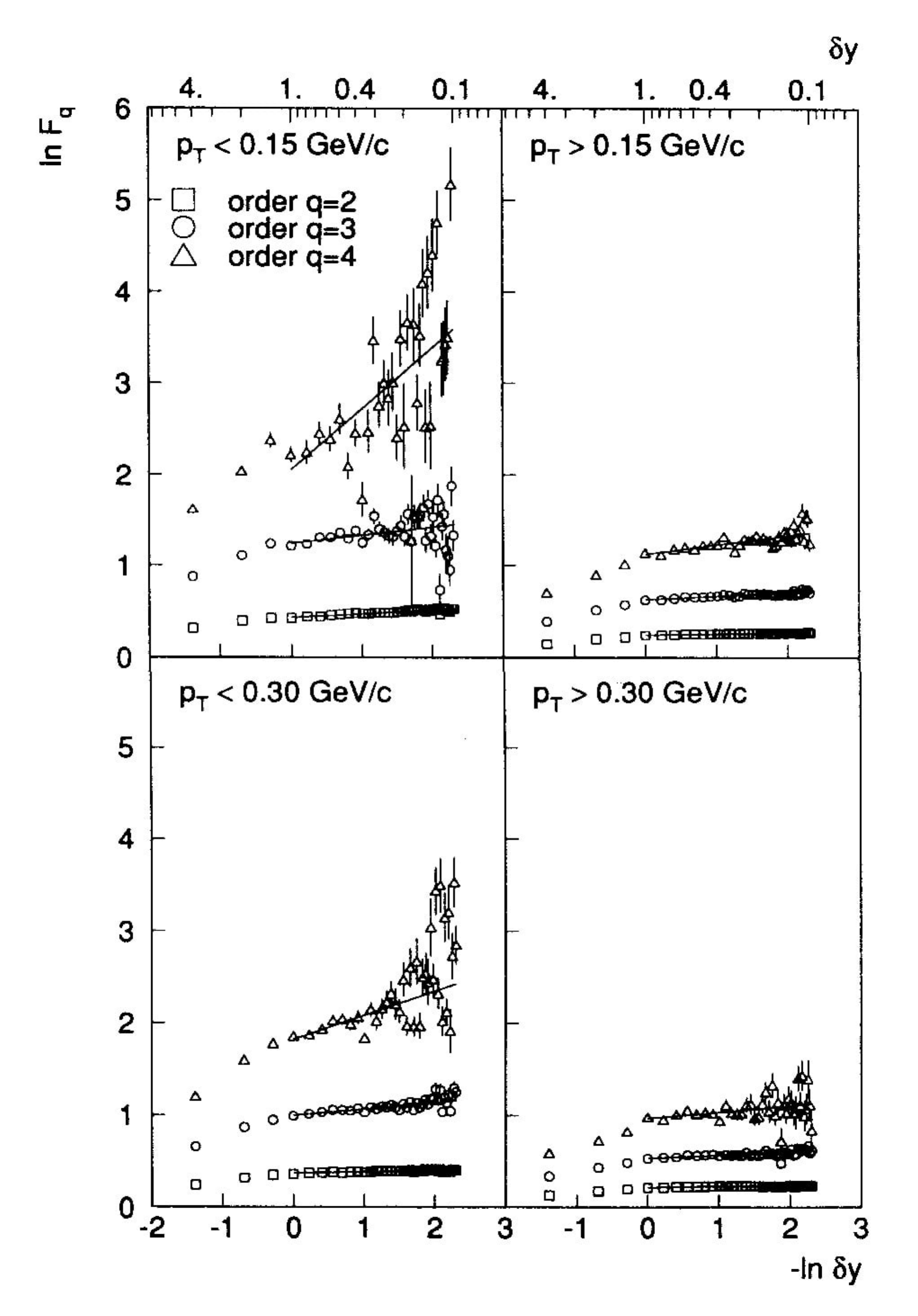}}
\caption{$\ln F_q$ as a function of $-\ln \d y$ for various $p_\rT$
cuts as indicated~\cite{Ajin89-90}.}
\end{figure} 

 UA1 has a bias against $p_\rT < 0.15$ GeV/$c$ and the anomalous
​ dimension is indeed smaller in UA1 than in NA22 in Fig. 15.
​ We conclude that intermittency in hh collisions is not dominated
 by semi-hard effects.

\subsection{Energy and multiplicity dependence}
As seen in Fig.~18, a strong multiplicity dependence of the intermittency
strength is observed for hh collisions by 
UA1~\cite{Alba90}. The trend is opposite to the predictions of the models 
used by this collaboration. This decrease of the intermittency strength with 
increasing multiplicity is usually explained as a consequence of mixing of 
independent sources of particles~\cite{LiBu89}. 

\begin{figure}[htb]
\includegraphics[width=6cm]{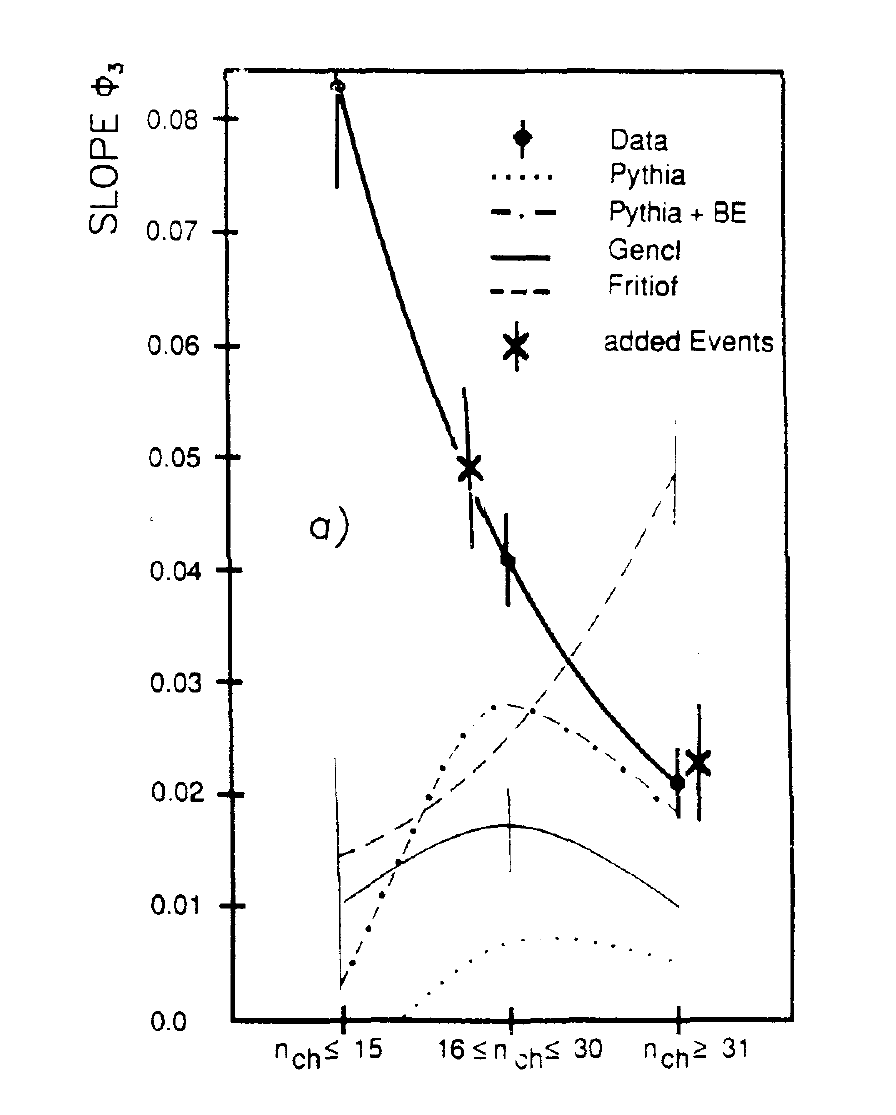}
\hs-3mm
\includegraphics[width=7cm]{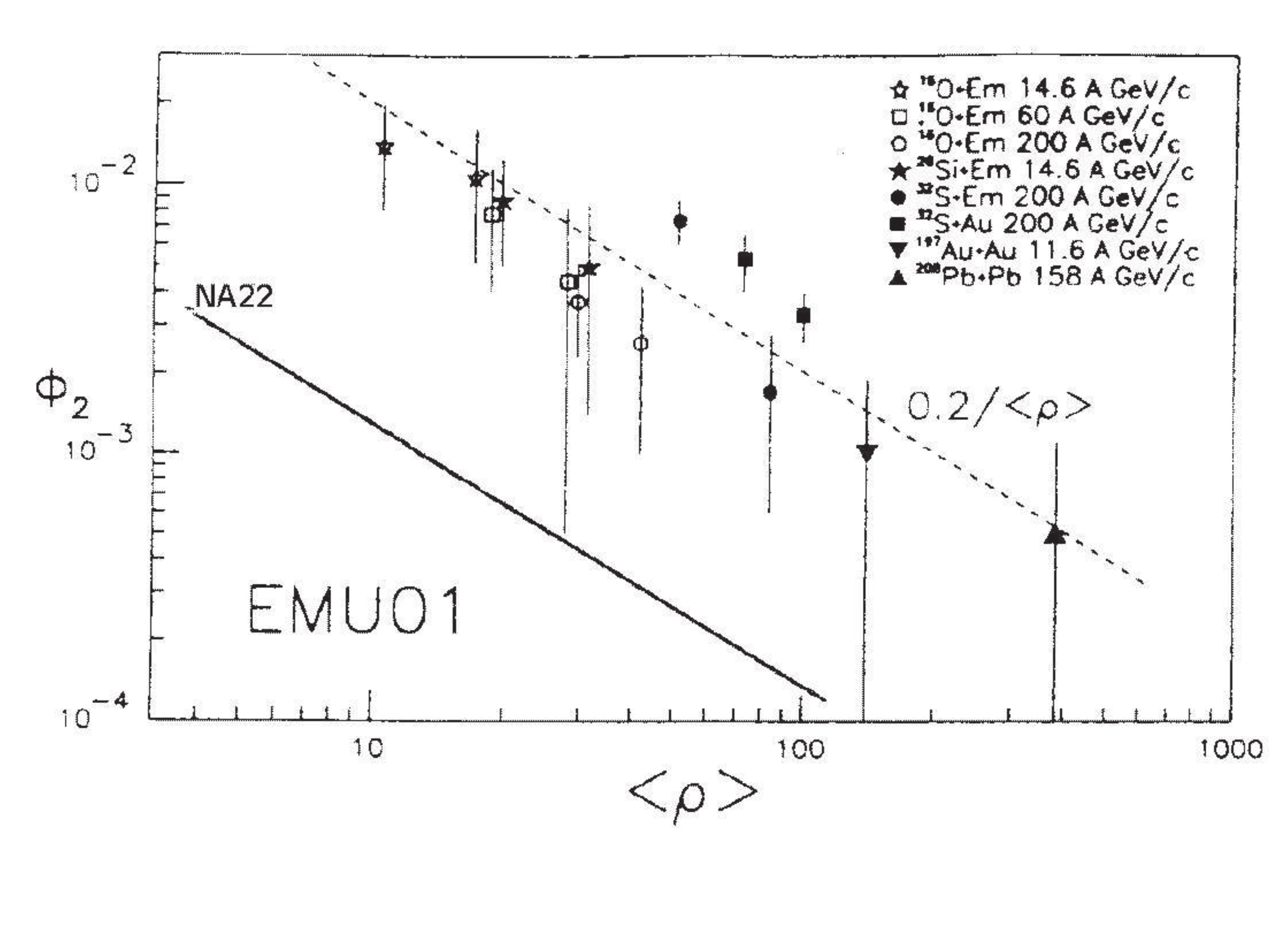}
\caption{ a) Multiplicity dependence of the slope 
$\f_3$, compared to that expected from a number of models, the crosses 
correspond to a combination of independent events~\cite{Alba90},
 b) slope $\f_2$ extrapolated $(\propto \r^{-1})$ as a function of particle 
density from NA22 (hp at 250 GeV) (solid line) and heavy-ion
collisions as indicated~\cite{Adamo90}.}
\end{figure}

Mixing of emission sources leads to a roughly linear 
decrease of the slopes $\f_q$ with increasing particle density $\lan \r\ran$ 
in rapidity~\cite{CapFia89,BiaFest89,Seib90}: $\f_q\propto \lan\r\ran^{-1}$. 
This is indeed observed by UA1~\cite{Alba90}.

Fig. 18a helps in explaining  why intermittency  is so weak in heavy-ion
collisions (cfr.~Fig.~15): the density (and 
mixing of sources) is particularly high there. In Fig.~18b, EMU01 
\cite{Adamo90}, therefore, compares $\f_2$ for NA22 (hp at 250 GeV) and 
heavy-ion collisions at similar beam momentum per nucleon, as a function of 
the particle density. Whereas slopes averaged over multiplicity are smaller 
for AA collisions than for NA22 in Fig.~15, at fixed 
$\lan \r\ran$ they are actually higher than expected from an extrapolation 
of hh collisions to high density and may even grow with 
increasing size of the nuclei. The trend is confirmed by KLMM \cite{cher1}
for intermittency in azimuthal angle $\vf$ and for 
slopes up to order 5.
This may be evidence for re-scattering (see~\cite{Verlu90}) or another 
(collective) effect, but, as shown by HELIOS 
\cite{AAke90} and confirmed by EMU-01~\cite{Adamo90}, one has to be very sure 
about the exclusion of $\g$-conversions before drawing definite conclusions.

\subsection{Density and correlation integrals}

\begin{figure}[b]
\centerline{
\includegraphics[width=9cm]{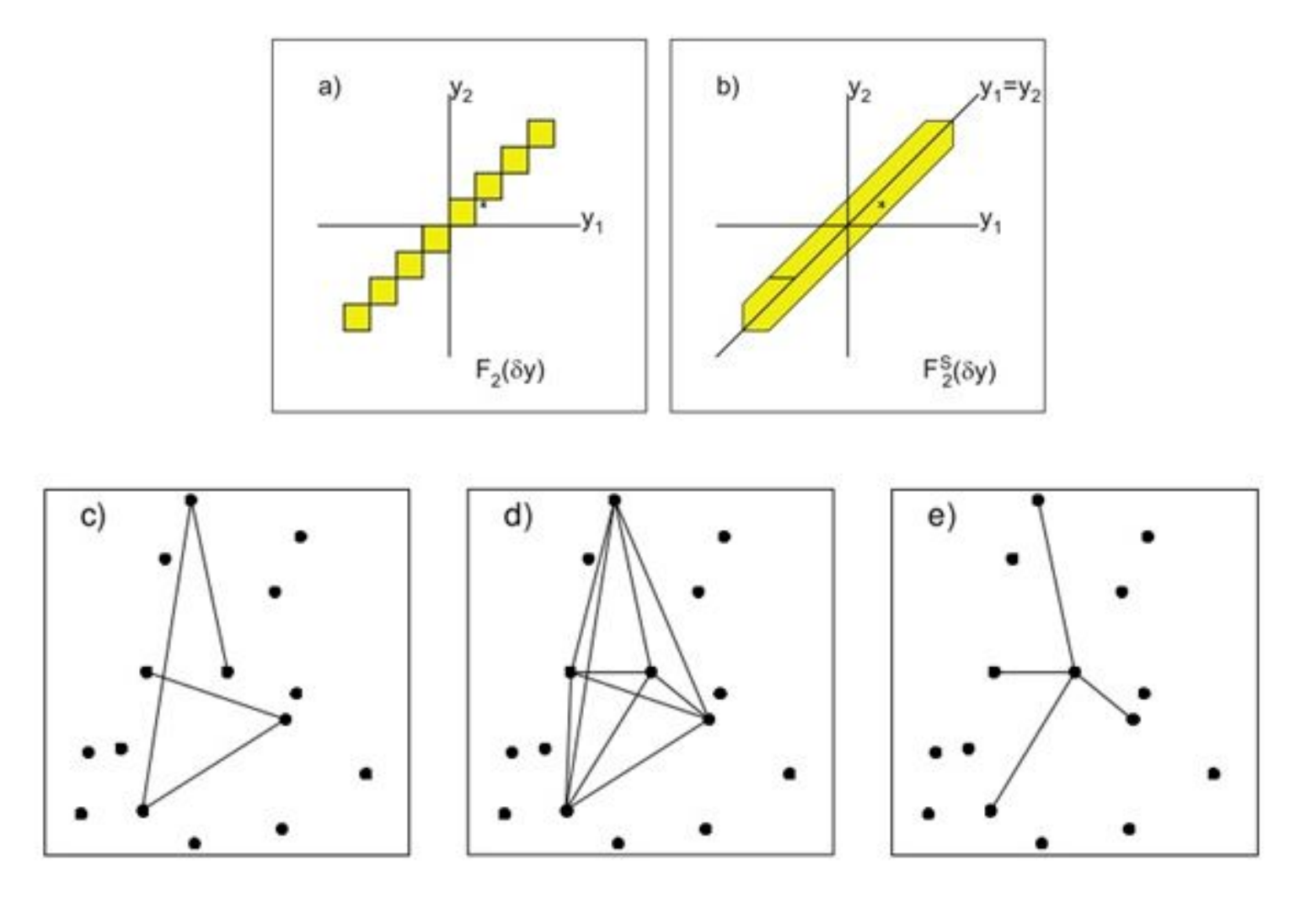}}
\caption{
a) The integration domain $\W_\rB=\S_m\W_m$ of $\r_2(y_1,y_2)$
for the bin-averaged factorial moments, b) the 
corresponding integration domain $\W_\rS$ for the density integral, 
c) illustration of a $q$-tuple in snake
topology, d) GHP topology, e) star topology \cite{Char94}.}
\end{figure} 

A fruitful development in the study of density fluctuations is 
the density and correlation strip-integral method \cite{Hen83} illustrated in Fig.~19 \cite{Char94}.
By means of 
integrals of the inclusive density over a strip domain in $y_1, y_2$ space, 
rather than a sum of box domains, one not only avoids unwanted side-effects 
such as splitting of density spikes, but also drastically increases the
integration volume (and therefore the statistical significance) at given 
resolution.
In terms of the strips (or hyper-tubes for $q>2$), the density 
integrals can be evaluated directly from the data after selection
of a proper distance measure, as e.g. the four-momentum difference 
$Q^2_{ij} = -(p_i-p_j)^2$,
and after definition of a proper multiparticle topology (snake integral \cite{CaSa89},
GHP integral \cite{Hen83}, star integral \cite{Egger93}).
Similarly, {\it correlation} integrals can be defined by replacing the
density $\r_q$ in the integral by the correlation function $C_q$.

\begin{figure}[htb]
\centerline{
\includegraphics[width=6cm]{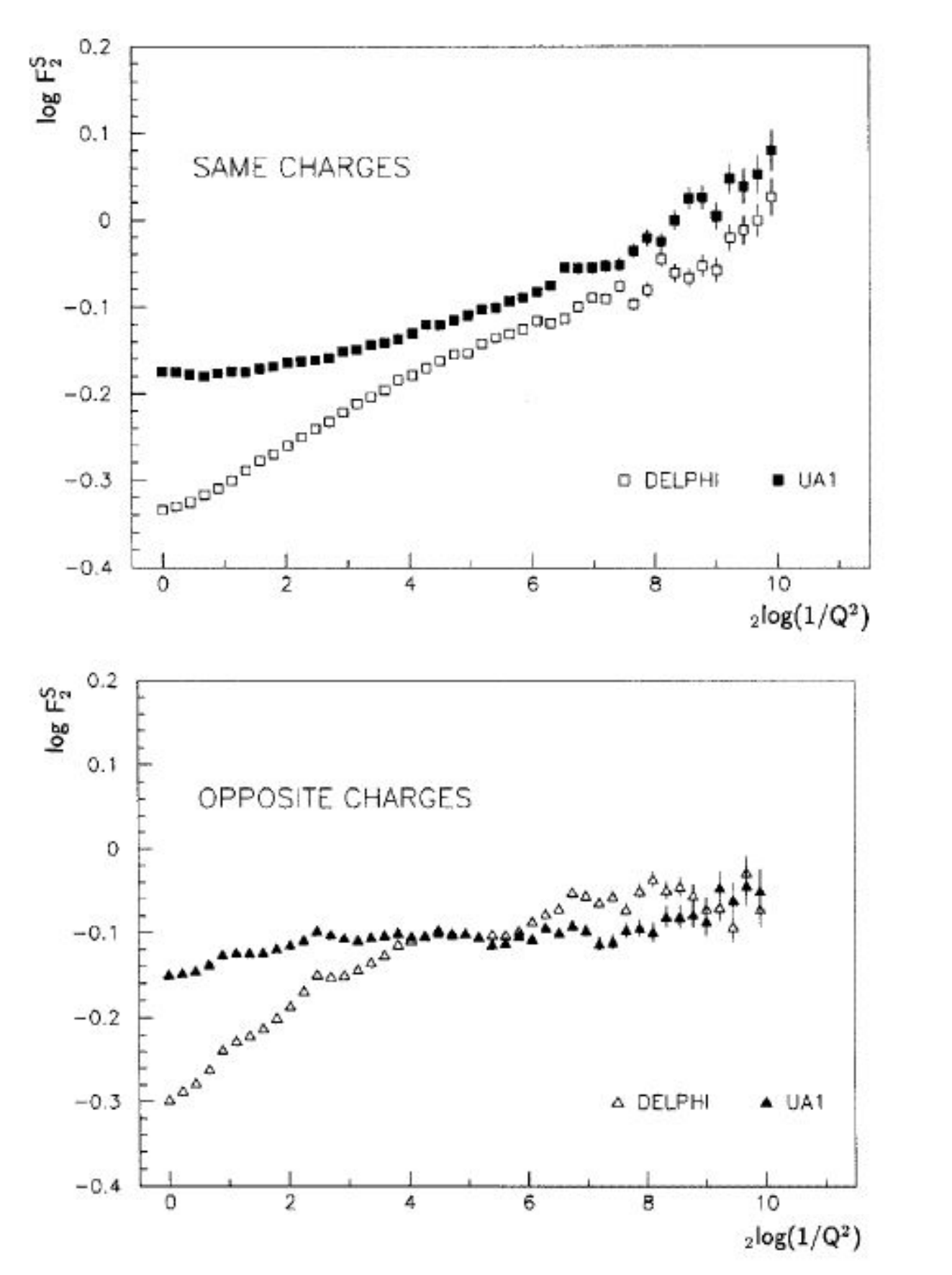}} 
\caption{
Comparison of density integrals for $q=2$ in
their differential form (in intervals $Q^2, Q^2+\rd Q^2$) 
as a function of $_2\log(1/Q^2)$ for e$^+$e$^-$ (DELPHI) 
and hadron-hadron collisions (UA1).\cite{Abreu94}}
\end{figure}

Of particular interest is a comparison of hadron-hadron to e$^+$e$^-$ results 
in terms of same and opposite charges of the particles involved. Such a 
comparison is 
shown in Fig.~20 for $q=2$ \cite{Abreu94}. An important difference between UA1 
and DELPHI can be observed in a comparison of the two sub-figures: 
For relatively large $Q^2(>0.03$ GeV$^2$), where Bose-Einstein effects do 
not play a major role, the e$^+$e$^-$ data increase much faster with 
increasing $-_2\log Q^2$ than the hadron-hadron results. For e$^+$e$^-$, 
the increase in this $Q^2$ region is very similar for same and for 
opposite-sign charges. 
At small $Q^2$, however, the e$^+$e$^-$ results approach the hh results. 
For e$^+$e$^-$ annihilation at LEP at least
two processes are considered to be responsible for the power-law behavior: Bose-Einstein
correlation at small $Q^2$ following the evolution of jets at larger $Q^2$ , but what is 
remarkable is the smooth transition between the two domains (if at all present) (see Sect. 3).

\subsection{Genuine higher-order correlations}

The correlation integral method turns out 
particularly useful for the unambiguous establishment of genuine higher-order 
correlations in terms of the normalized cumulants
$K_q(Q^2)$, when using the star integration \cite{Egger93}.

Non-zero values of (star integral) $K^*_q(Q^2)$ increasing according to a power 
law with decreasing $Q^2$ were first observed in NA22 up to
fifth order \cite{genuine} (see Fig.~21) and in E665 for third order \cite{Adams94}. 
Again note the difference between all charged and like charged particles
​and the smooth transition betwen larger and smaller $Q^2$.

\begin{figure}[htb]
\centerline{
\includegraphics[width=8cm]{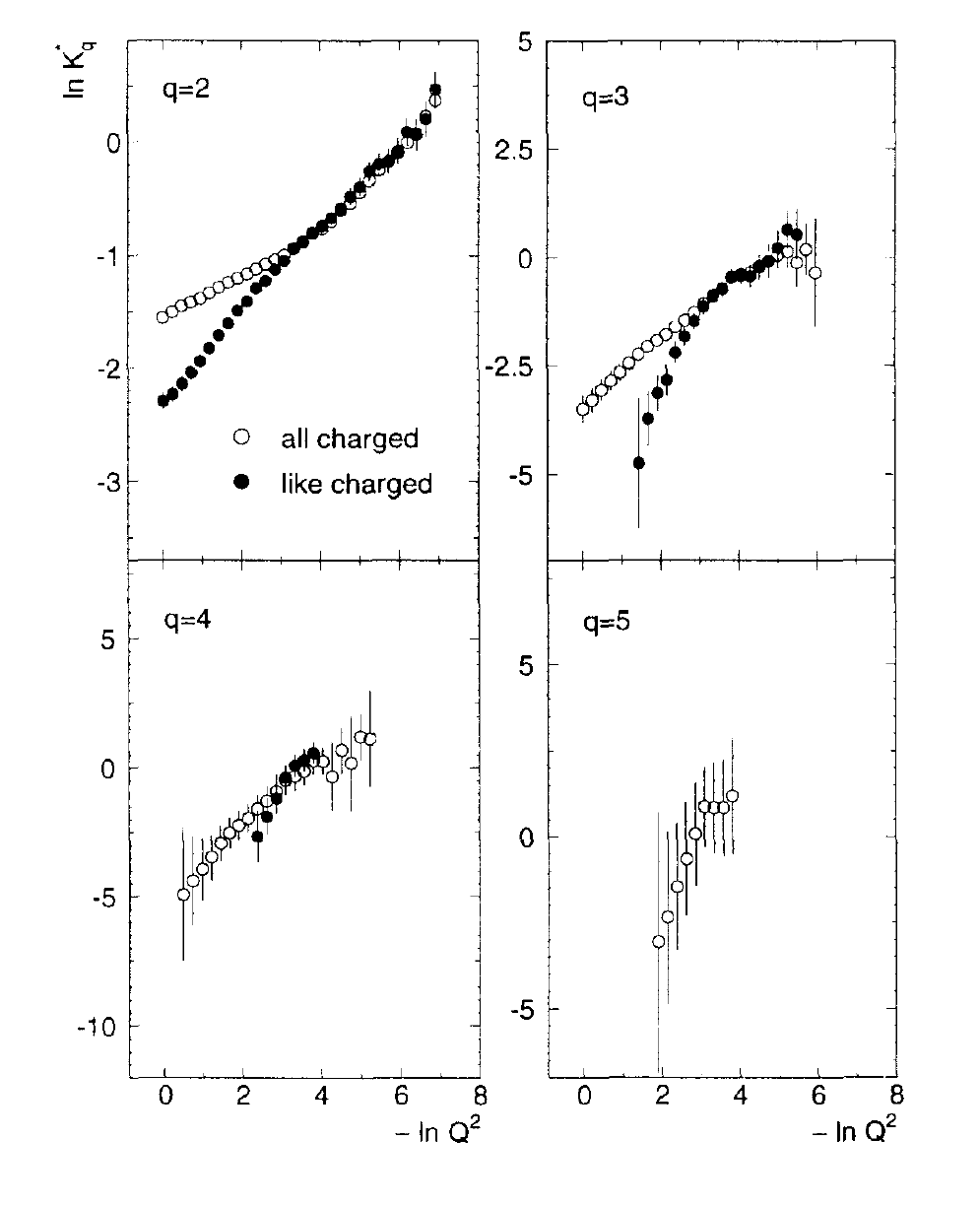}}
\caption{
 $\ln K^*_q(Q^2)$ as a function of $-\ln Q^2$ for all charged
particles as well as for like-charged particles \cite{genuine}.} 
\end{figure}

 \subsection{Functional form}
The exact functional form of $F^\rS_2$ is derived from the data of 
UA1~\cite{Alba90} and NA22~\cite{Char94}\footnote{In fact in this 
form $F^\rS_2(Q^2)$ is identical 
to $R(Q^2)$ usually used in Bose-Einstein analysis. The only difference is 
that it is plotted on a double-logarithmic plot, here.} in Fig.~22. 
Clearly, the data favour a power law in $Q$  over an 
exponential, double-exponential or Gaussian law.

\begin{figure}[t]
\centerline{
\includegraphics[width=6cm]{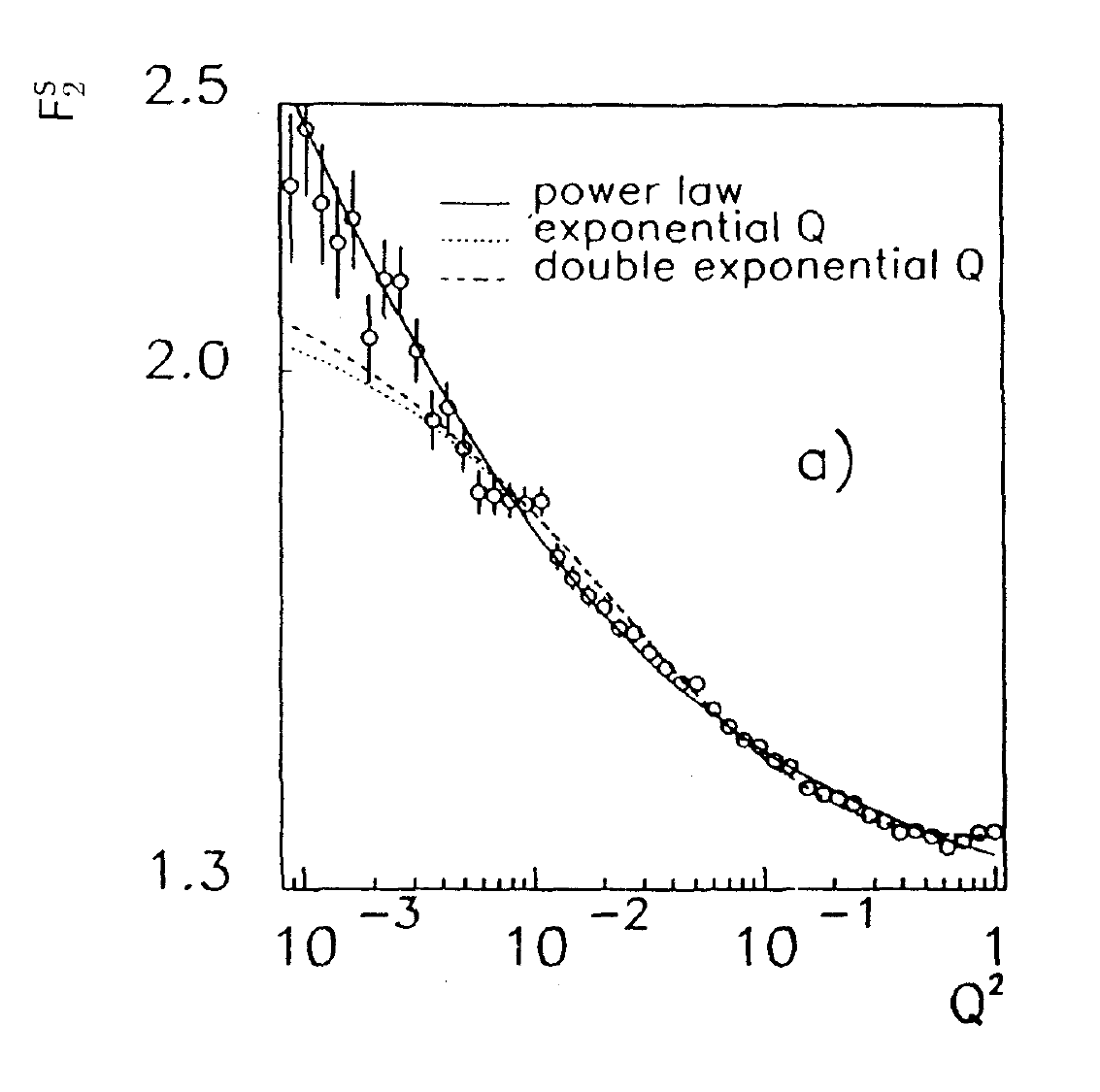}\hskip1mm
\includegraphics[width=6cm]{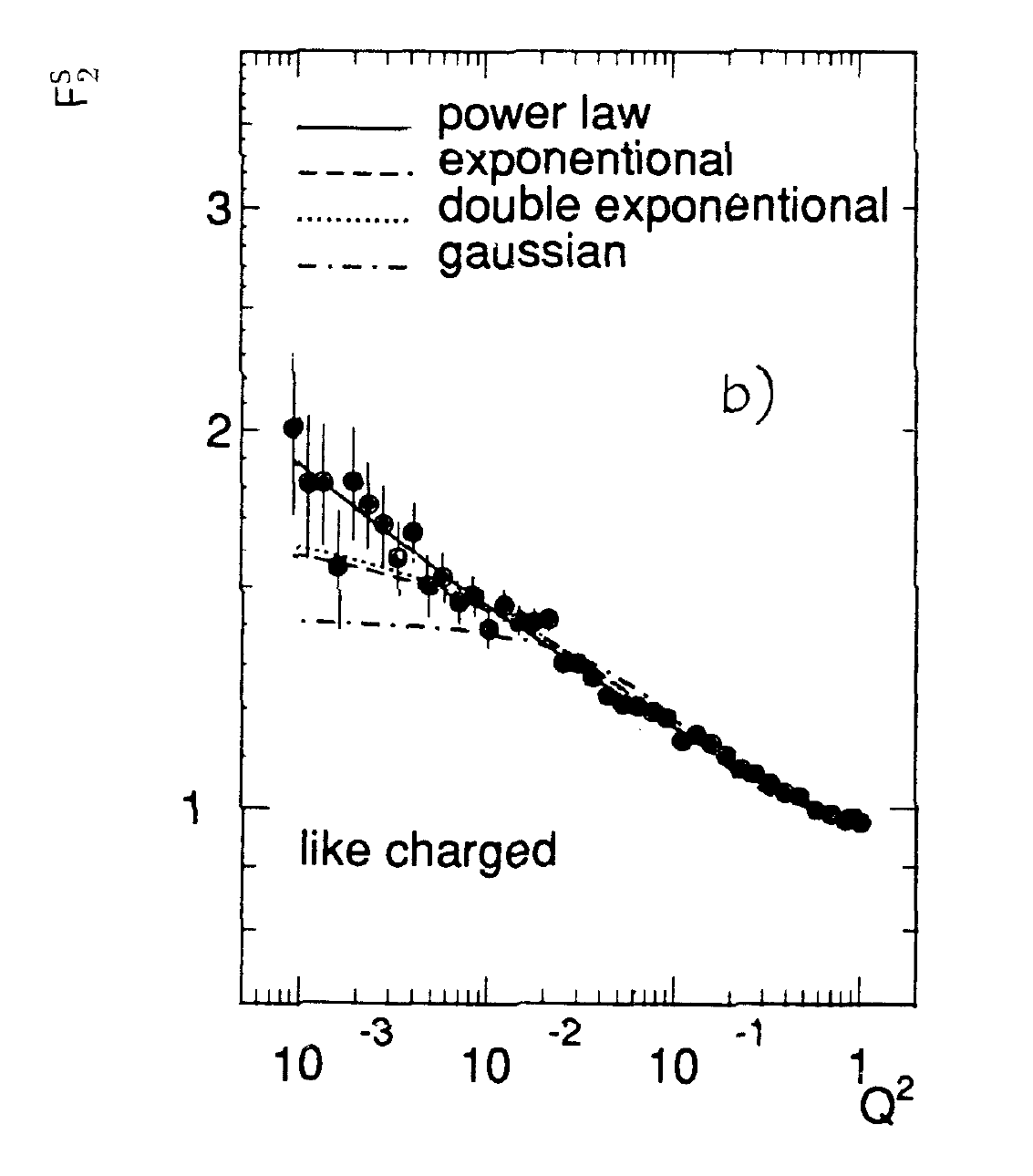}}
\caption{Density integrals $F^\rS_2$ (in 
their differential form) as a function of $Q^2$ for like-charged 
pairs in UA1~\cite{Alba90} and NA22~\cite{Char94}, 
compared to power-law, exponential,
double-exponential and Gaussian fits, as indicated. } 
\end{figure}

If the observed effect is real, it supports a view developed
in~\cite{Bial92}. There, intermittency is explained from 
Bose-Einstein correlations between 
(like-charged) pions. As such,  Bose-Einstein correlations from a static 
source\index{source!static} are not power behaved. A power law
is obtained i) if the size of the interaction region is allowed to fluctuate, 
and/or ii) if the interaction region itself is assumed to be a 
self-similar object extending over a large volume. 
Condition ii) would be realized if parton avalanches were to arrange 
themselves into self-organized critical states \cite{Bak87}. 
Though quite speculative at this moment, it is an 
interesting new idea with possibly far-reaching implications. 
We should mention also that in such a scheme intermittency is viewed as 
a final-state effect and is, therefore, not troubled by 
hadronization effects. 

 So, in conclusion of this section, (approximate) intermittency
​ is found to be all-present in hadron production and is evidence
​ for genuine correlations to high orders, but it seems dominated by
​ Bose-Einstein correlations. However, what we have learned
​ is that we have been fooled for more than half a century by
 an assumed Gaussian behavior of the BE correlations, while
​ an approximate power law is required. This highly non-trivial
​ lesson we have learned indeed throws a completely new
​ light on the topic of femtoscopy.

\section{Bose-Einstein Correlations (or what?)}
\subsection{Early results}
Whether derived as Fourier transform of a (static and chaotic) 
pion source distribution, a covariant Wigner-transform of the
(momentum dependent) source density matrix, or from the string model, 
identical-pion correlation leads to a positive, non-zero two-particle 
correlator $K_2(Q)$, i.e. to
\beq
R_2(Q)=1+K_2(Q)>1
\eeq
at small four-momentum difference $Q$.
These so-called Bose-Einstein Correlations, by now, are a well-established
effect in all types of collisions,
even in hadronic Z$^0$ decay (for reviews see \cite{kitwolf2005,Kit01,Ale03})
originally expected, however, to be too coherent to show an effect.

Other important observations are given in abstract form below.

1. When evaluated in two (or better three) dimensions in the Bertsch-Pratt
system, a small elongation of the emission region (better region of
homogeneity \cite{Sin} is observed along the event axis in all types of
collisions (hadron-hadron \cite{NA22}, all four LEP experiments \cite{LEPEL},
ZEUS \cite{ZEUS}, RHIC \cite{Mag}).
However, it is important to note that the longitudinal radius of
homogeneity is much shorter than the length of the sting (of order 1\%).

The observation that the out-radius does not grow beyond 
the side-radius at RHIC \cite{Mag} points
to a short duration of emission and 
causes a problem for some hydrodynamical models, but not for
e.g. the Buda-Lund hydro model. The latter, in fact gives a beautifully
consistent description of single-particle spectra and BEC in hadron-hadron
and heavy-ion collisions at SPS and RHIC \cite{Ster}.
The emission function resembles a Gaussian shaped fire-ball for AA 
collisions, but a fire-tube for hh collisions.

2. The form of the correlator at small $Q$ is steeper than Gaussian,
in fact consistent with a power law as would be expected from the
intermittency phenomenon described above. Unifying progress is
reported in \cite{Cso}.

3.  An approximate $m^{-1/2}_\rT$ scaling first observed in heavy-ion
collisions at the SPS \cite{SPS} and usually blamed on collective flow, is
observed at RHIC \cite{PHEN-PHO}, but also in e$^+$e$^-$ collisions \cite{MT}.
Quite generally, it follows from a strong position momentum correlation
\cite{Bia}, be it due to collective flow or to string fragmentation.

4. {\em Genuine three-pion correlations} exist in all types of collisions
and, in principle, allow a phase to be extracted from
\beq
\cos\f \equiv \w(Q_3)= K_3(Q_3)/2\sqrt{K_2(Q_3)} .
\eeq
At small $Q$, this $\w$ is near unity (as expected from incoherence)
for hh \cite{NA22-2} and e$^+$e$^-$ \cite{L3} collisions, as well as for
PbPb \cite{NA44,WA98} and AuAu \cite{STAR03} collisions at SPS and RHIC, while
it is near zero (compatible with full coherence) in collisions of
light nuclei \cite{NA44}. This contradiction can be solved \cite{Kit01,Br}
if $\w$ is interpreted as a ratio of normalized cumulants.
Since $K^{(N)}_q$ of $N$ independent overlapping sources gets diluted
like $1/N^{q-1}$, $\w$ would be reduced if strings produced by light ions
do not interact. If, in {\em heavy} ion collisions,
the string density gets high enough for them to coalesce, some kind
of percolation sets in and full inter-string BEC gets restored.

5. Azimuthal anisotropy is observed in configuration space of
non-central heavy-ion collisions at AGS energies \cite{ANIS}, but
also at RHIC \cite{STAR2}. Contrary to elliptic flow, it is directed out
of the event plane, but consistent with the elliptic nuclear
overlap in a non-central collision. Due to larger pressure in the event
plane, the anisotropy gets reduced but not destroyed at RHIC.
Also this is evidence for a short duration of pion emission.

\subsection{The $\tau$ model} 

In e$^+$e$^-$, BEC depend, at least approximately, only on $Q$ and not on its components separately, in 
the sense that e$^+$e$^-$  BEC is large if Q is small even when any of its components are large.
Further, $R_2$ shows anti-correlations in the region 0.6--1.5 GeV as observed by L3 at LEP \cite{L3-Achard}
as well as by CMS \cite{CMS2011} and ATLAS (preliminary) \cite{Asta} at LHC (see Fig. 23).  
 
 A model which predicts such $Q$-dependence, as well as the absence of dependence on the components of $Q$ separately,
is the so-called $\tau$-model~\cite{Tamas;Zimanji:1990}.
Further it incorporates the  Bjorken-Gottfried condition \cite{Bialas:1999} whereby
the four-momentum of a produced particle and the space-time position at which it is produced are linearly
related.
 
\begin{figure}
\begin{minipage}{7cm}
\includegraphics[width=6.9cm]{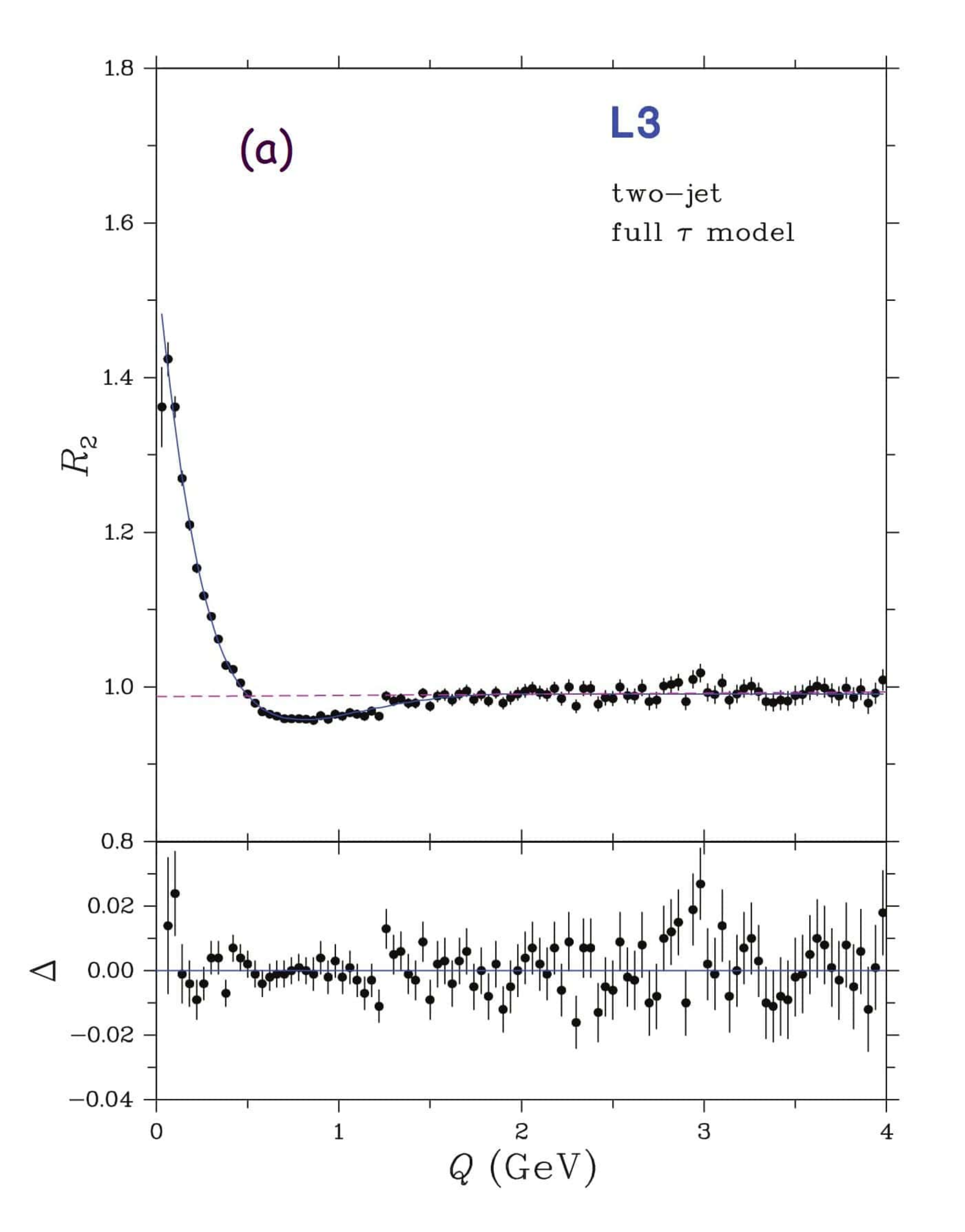} 
\end{minipage}
\hskip-3mm
\begin{minipage}{6cm}\vskip-2mm
\includegraphics[width=6cm]{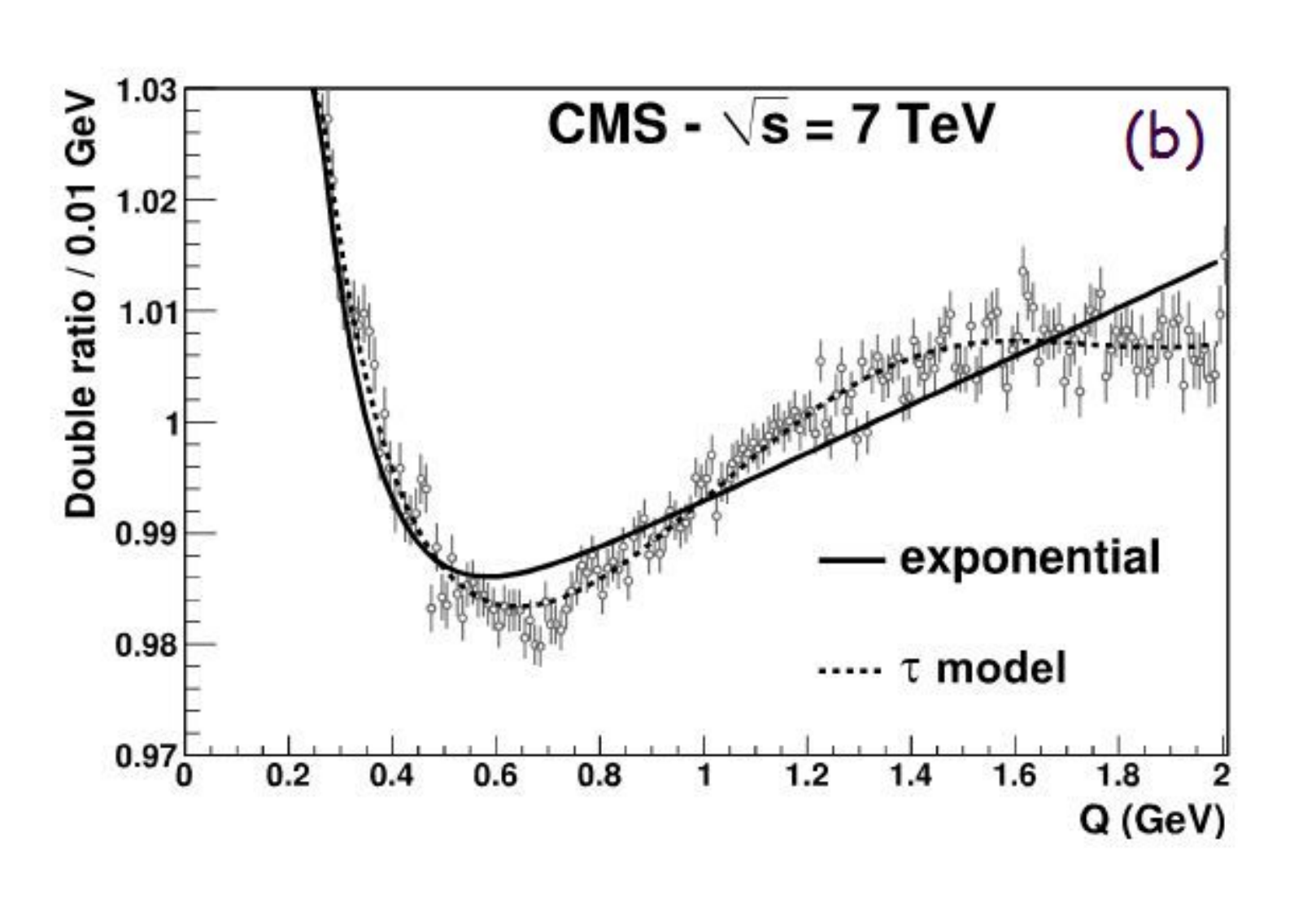}\vskip-2mm\hskip-1mm
\includegraphics[width=6cm]{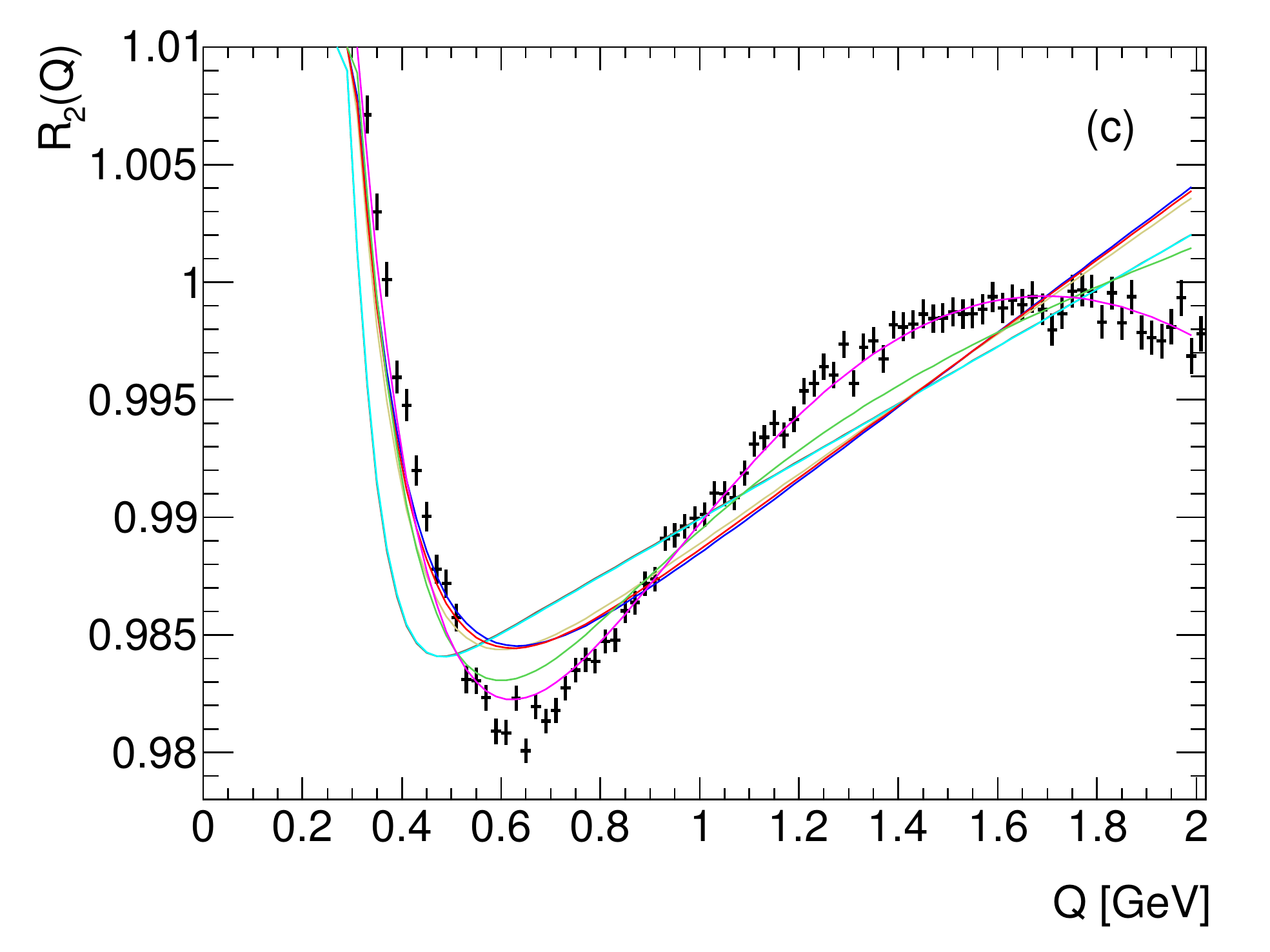} 
\end{minipage}
  \caption{The Bose-Einstein correlation function $R_2$ for a) L3 \cite{L3-Achard},  b) CMS \cite{CMS2011} 
and c) ATLAS \cite{Asta}. 
The curve in (a), the dashed line in (b) and the best fit in (c)
correspond to the fit of the $\tau$ model. 
           The results of the L3 fit are given in Table 1.
           Also plotted in (a) is $\Delta$, the difference between the fit and the data.
           The dashed line represents the long-range part of the fit, i.e., $\gamma(1+\epsilon Q)$.
The full line in (b) is an exponential fit. The lines in (c) correspond to Gaussian, exponential, and $\tau$ model fits.}
\end{figure}

In this model, it is assumed that the average production point in the overall center-of-mass system,
$\overline{x}=(\overline{t},\overline{r}_\mathrm{x},\overline{r}_\mathrm{y},\overline{r}_\mathrm{z})$,
of particles with a given four-momentum $p=(E,\px,\py,\pz)$ is given by
\begin{equation} \label{eq:tau-corr}
   \overline{x}^\mu (p^\mu)  = a\tau p^\mu \;.
\end{equation}
In the case of two-jet events,
  $a=1/m_\rT$
where
$m_\rT$ is the transverse mass
and
$\tau = \sqrt{\overline{t}^2 - \overline{r}_{\kern -0.14em \mathrm{z}}^2}$ is the longitudinal proper time.
For isotropically distributed particle production, the transverse mass is replaced by the
mass in the definition of $a$ and $\tau$ is the proper time,
$\sqrt{\overline{t}^2 - \overline{r}_{\kern -0.14em \mathrm{x}}^2
                      - \overline{r}_{\kern -0.14em \mathrm{y}}^2
                      - \overline{r}_{\kern -0.14em \mathrm{z}}^2}$.
 
The second assumption is that the distribution of $x^\mu (p^\mu)$ about its average,
$\delta_\Delta ( x^\mu(p^\mu) - \overline{x}^\mu (p^\mu) )$, is narrower than the
proper-time distribution, $H(\tau)$.
Then   the two-particle Bose-Einstein correlation function is indeed found to depend on the invariant
relative momentum $Q$, rather than on its separate components, as well as on the values of $a$ of the two particles \cite{ourTauModel}:
\begin{equation} \label{eq:levyR2}
   R_2(p_1,p_2) = 1 +         \mathrm{Re} \widetilde{H}\left(\frac{a_1 Q^2}{2}\right)
                                          \widetilde{H}\left(\frac{a_2 Q^2}{2}\right) \;,
\end{equation}
where $\widetilde{H}(\omega) = \int \mathrm{d} \tau H(\tau) \exp(i \omega \tau)$
is the Fourier transform (characteristic function) of $H(\tau)$.
(Note that $H(\tau)$ is normalized to unity.)

Since there is no particle production before the onset of the collision,
$H(\tau)$ should be a  one-sided distribution.
In the leading log approximation of QCD  the parton shower
is a fractal~\cite{Dahlqvist:1989yc}.
Further,  a L\'evy distribution arises naturally from a fractal~\cite{metzler}.
One is thus led to choose a one-sided L\'evy distribution for $H(\tau)$~\cite{ourTauModel}.
The characteristic function of $H(\tau)$ can then be written~\cite{Tamas:Levy2004}  (for $\alpha\ne1$)
as

\begin{equation} \label{eq:levy1sidecharf}
   \widetilde{H}(\omega) = \exp\left[ -\frac{1}{2}\left(\Delta\tau|\omega|\strut\right)^\alpha
          \left( 1 -  i\, \mathrm{sign}(\omega) \tan\left(\frac{\alpha\pi}{2}\right) \strut \right)
       + i\,\omega\tau_0\right]
 \; ,
\end{equation}
where the parameter $\tau_0$ is the proper time of the onset of particle production
and $\Delta \tau$ is a measure of the width of the proper-time distribution.
$0<\a<2$ is the so-called index of stability \cite{Nolan:2010} of the L\'evy distribution. 
Using this characteristic function in (17),
and incorporating the usual strength factor $\lambda$ and the long-range parametrization,
yields
\begin{equation} 
\begin{split}
   R_2(Q,a_1,a_2) &= {} \gamma \left\{  1 +
      \lambda\cos\left[\frac{\tau_0 Q^2 (a_1+a_2)}{2} +
\tan\left(\frac{\alpha\pi}{2}\right)\left(\frac{\Delta\tau {Q^2}}{2}\right)^{\!\alpha}\frac{a_1^\alpha+a_2^\alpha}{2} \right]
  \right.
\\ 
     &
  \left.
     \quad \cdot           \exp \left[-\left(\frac{\Delta\tau {Q^2}}{2}\right)^{\!\alpha}\frac{a_1^\alpha+a_2^\alpha}{2} \right]
       \right\} \left(1+\epsilon Q\right)
 \; .
\end{split}
\end{equation}
 
It is the cosine factor which generates oscillations corresponding to alternating correlated and anti-correlated regions mentioned above.
Note also that since $a=1/m_\rT$ for two-jet events, the \taumodel\ predicts a decrease of the effective source size with increasing $m_\rT$.  

\begin{table}[h]
\caption{Results of the fit of (19)\ for two-jet events,
         as shown in Fig.~23 a) \cite{L3-Achard}.
         The parameter $\tau_0$ is fixed to zero.
         The first uncertainty is statistical, the second systematic.}
\begin{center}
\renewcommand{\arraystretch}{1.2}
\begin{tabular}{ l r@{$\;\pm\;$}l
               }
\hline
    parameter               & \multicolumn{2}{c }{\ }           \\
\hline
  $\lambda$                 & 0.58  & $0.03^{+0.08}_{-0.24}$    \\
  \rule{0pt}{11pt}$\alpha$  & 0.47  & $0.01^{+0.04}_{-0.02}$    \\
  $\Delta\tau$   (fm)       & 1.56  & $0.12^{+0.32}_{-0.45}$    \\
  $\epsilon$ (GeV${-2}$)        & 0.001 & $0.001\pm0.003$           \\
  $\gamma$                  & 0.988 & $0.002^{+0.006}_{-0.002}$ \\
\hline
  \chisq/DoF                & \multicolumn{2}{c }{90/95}        \\
  confidence level          & \multicolumn{2}{c }{62\%}         \\
\hline
\end{tabular} 
\end{center}
\end{table}
 
 For each bin in $Q$ the average values of $m_{\mathrm{T}1}$ and $m_{\mathrm{T}2}$ are calculated,
where $m_{\mathrm{T}1}$ and $m_{\mathrm{T}2}$ are the transverse masses of the
two particles making up a pair, requiring  $m_{\mathrm{T}1} > m_{\mathrm{T}2}$.
Using these averages, (19) is fit to $R_2(Q)$ by the L3 Coll. \cite{L3-Achard} .
The fit results in $\tau_0=0.00\pm0.02$~fm, and the results of a re-fit with $\tau_0$ fixed to zero
 are shown in Table 1.

Note that no significant long-range correlation is observed: $\epsilon=0$ well within one
 standard deviation and $\gamma$ is close to unity. Obviously, the $\tau$-model by itself 
can reproduce the (smooth) shape of the $Q$-distribution over the full range considered, the anticorrelation near 
$Q$=0.6 GeV included.

In the $\tau$-model, the basic assumption is the Bjorken-Gottfried condition \cite{Bialas:1999}
leading to (16). Recently, it has been demonstrated by the same authors \cite{bia;zal2013}, however, that already the 
compositeness of pions can most naturally lead to an anti-correlation. At small distances the constituents
mix and  there are no separate pions to interfere.

\subsection{The emission function}  
The $\tau$ model results for BEC can be used together with the single-particle inclusive
​spectra to reconstruct the space-time evolution of hadronization. The emission function in 
configuration space, $S_\mathrm{x}(x)$, is the proper time derivative of the
integral over $p$ of $S(x,p)$ \cite{ourTauModel}.
Approximating $\delta_\Delta$ by a Dirac delta function again, gives
\begin{equation}   \label{eq:Sspace}
   S_\mathrm{x}(x) = \frac{1}{\bar{n}} \frac{\mathrm{d}^4 n}{\mathrm{d}\tau\mathrm{d}^{3}x}
                   = \left(\frac{m_\rT}{\tau}\right)^3 H(\tau) \rho_1\left( p=\frac{m_\rT x}{\tau} \right) \;,
\end{equation}
where $n$ and $\bar{n}$ are the number and average number of pions produced, respectively, 
and $\rho_1(p)$ is the experimentally measurable single-particle spectrum.
 
Given the symmetry of two-jet events, $S_\mathrm{x}$ does not depend on the azimuthal angle, and one can
write it in cylindrical coordinates as
\begin{equation}   \label{eq:Srzt}
     S_\mathrm{x}(r,z,t) = P(r,\eta) H(\tau) \;,
\end{equation}
where $\eta$ is the space-time rapidity.
With the strongly correlated phase-space of the $\tau$-model,
$\eta$  is equal to the momentum-energy rapidity $y$ and $r=\pT\tau/\mT$.
Consequently,
\begin{equation}  \label{eq:Preta}
     P(r,\eta) = \left(\frac{\mT}{\tau}\right)^{\!3} \rho_\mathrm{\pT,y}(r\mT/\tau, \eta) \;,
\end{equation}
where $\rho_\mathrm{\pT,y}$ is the joint single-particle distribution of \pT\ and $y$.
 
The reconstruction of $S_\mathrm{x}$ is simplified if $\rho_\mathrm{\pT,y}$
can be factorized into the product of the single-particle \pT\ and rapidity distributions, \ie,
$     \rho_\mathrm{\pT,y} = \rho_\mathrm{\pT}(\pT) \rho_\mathrm{y}(y)$.
Then (22) becomes
\begin{equation}   \label{eq:fact}
     P(r,\eta) = \left(\frac{\mT}{\tau}\right)^{\!3} \rho_\mathrm{\pT}(r\mT/\tau) \rho_\mathrm{y}(\eta) \;,
\end{equation}

\begin{figure}[htb]
\centerline{
\includegraphics[width=8cm]{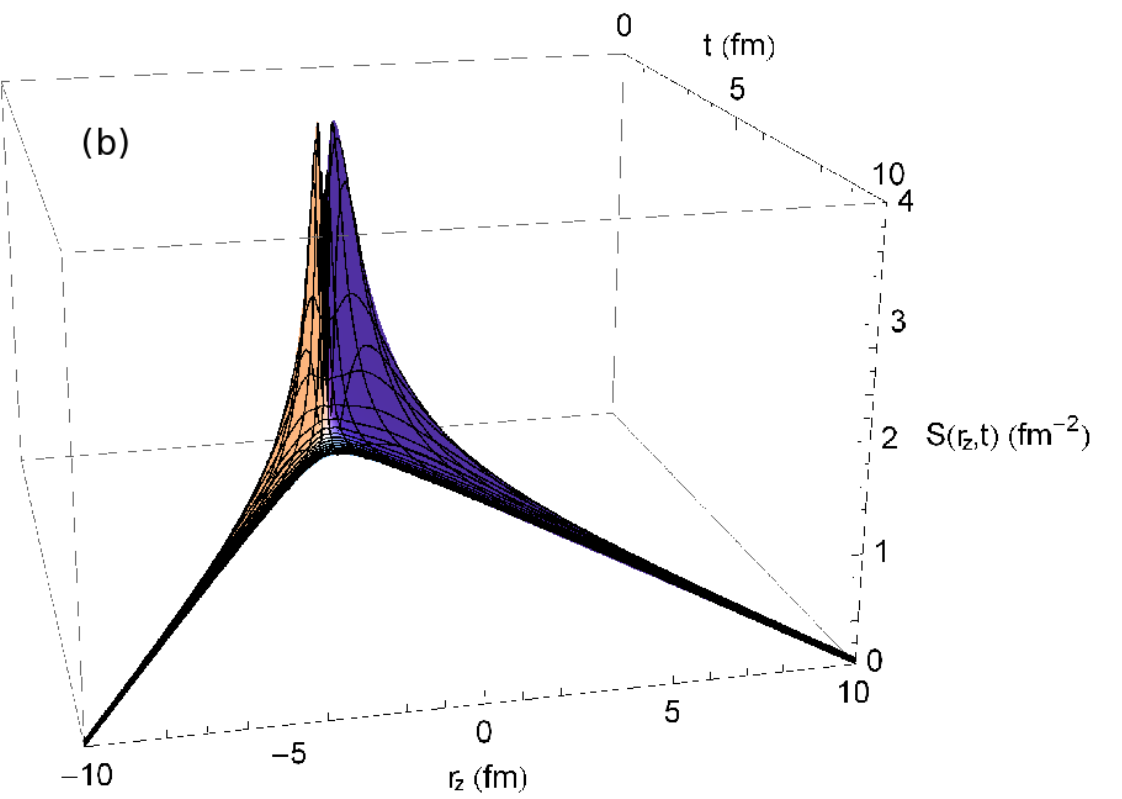} 
}
\caption{The temporal-longitudinal part of the emission function normalized to unity \cite{L3-Achard}.}
\end{figure}

\begin{figure}[htb]
\centerline{
\includegraphics[width=12cm]{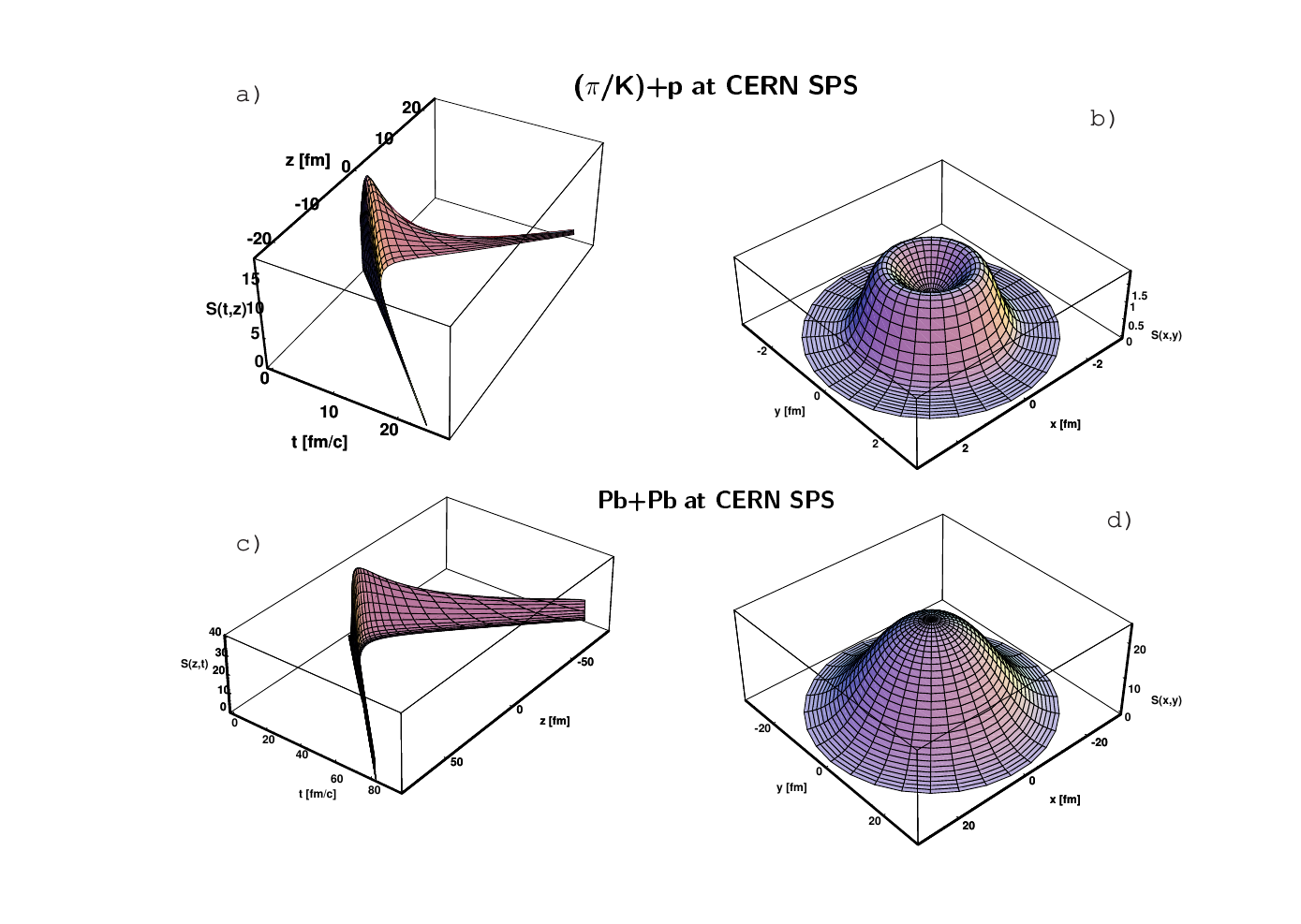}
}
\caption{
The reconstructed emission function
$S(t,z)$ in arbitrary vertical units, as a function of time $t$ and longitudinal 
coordinate $z$ (left diagrams), as well as the reconstructed emission 
function $S(x,y)$ in arbitrary vertical units, as a function of the 
transverse coordinates $x$ and $y$ (right pictures), for hh
 (upper pictures) and PbPb (lower pictures) collisions, respectively \cite{Agab1998,Ster2006,Ster1999}.}
\end{figure}

\begin{figure}[htb]
\includegraphics[width=5cm]{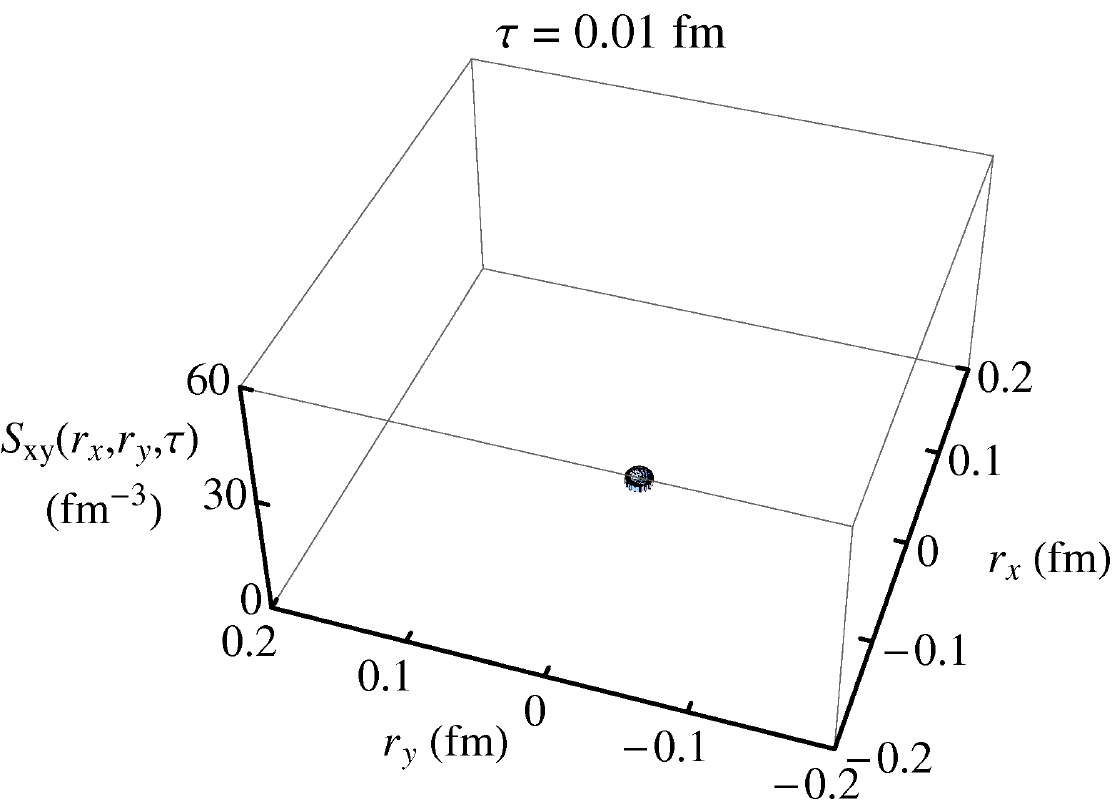} \includegraphics[width=5cm]{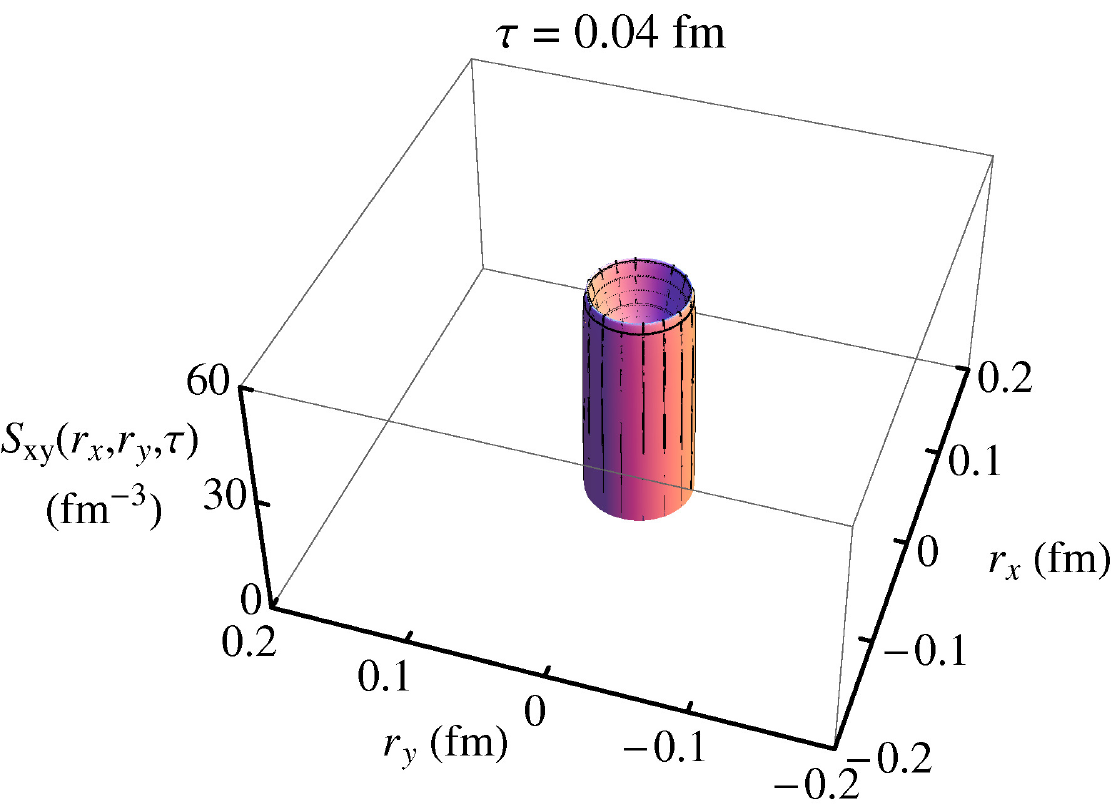} \\
\includegraphics[width=5cm]{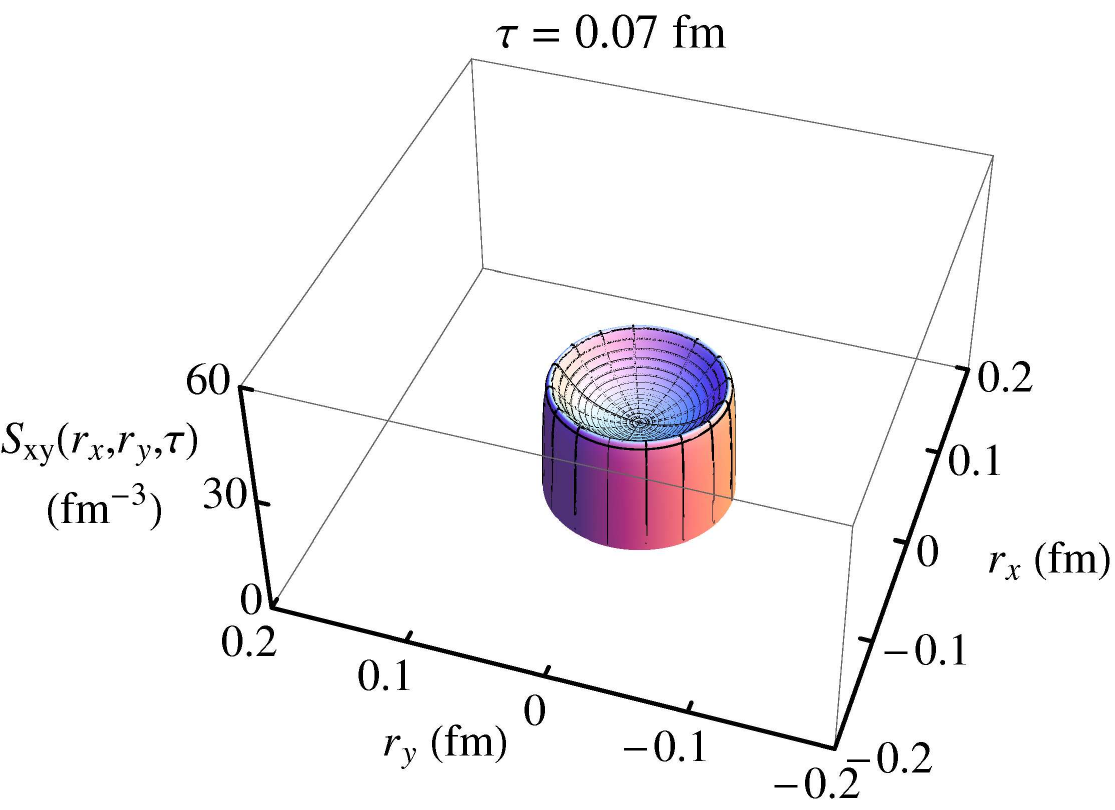} \includegraphics[width=5cm]{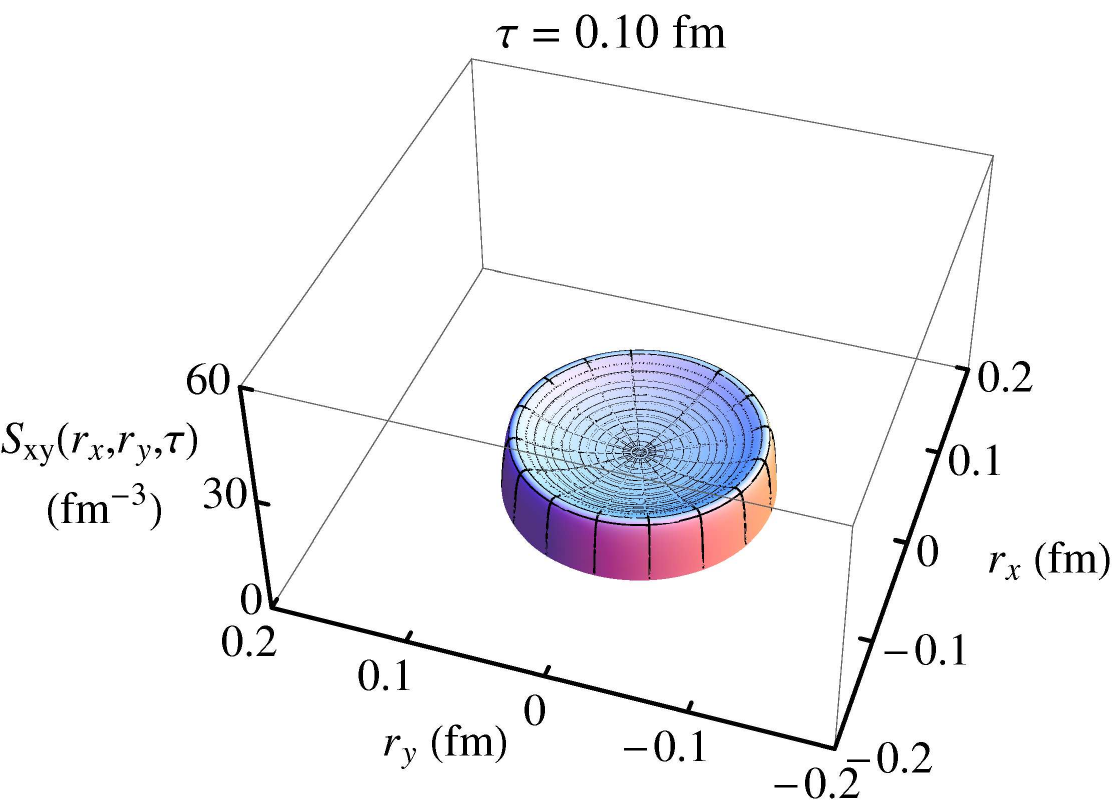} \\
\includegraphics[width=5cm]{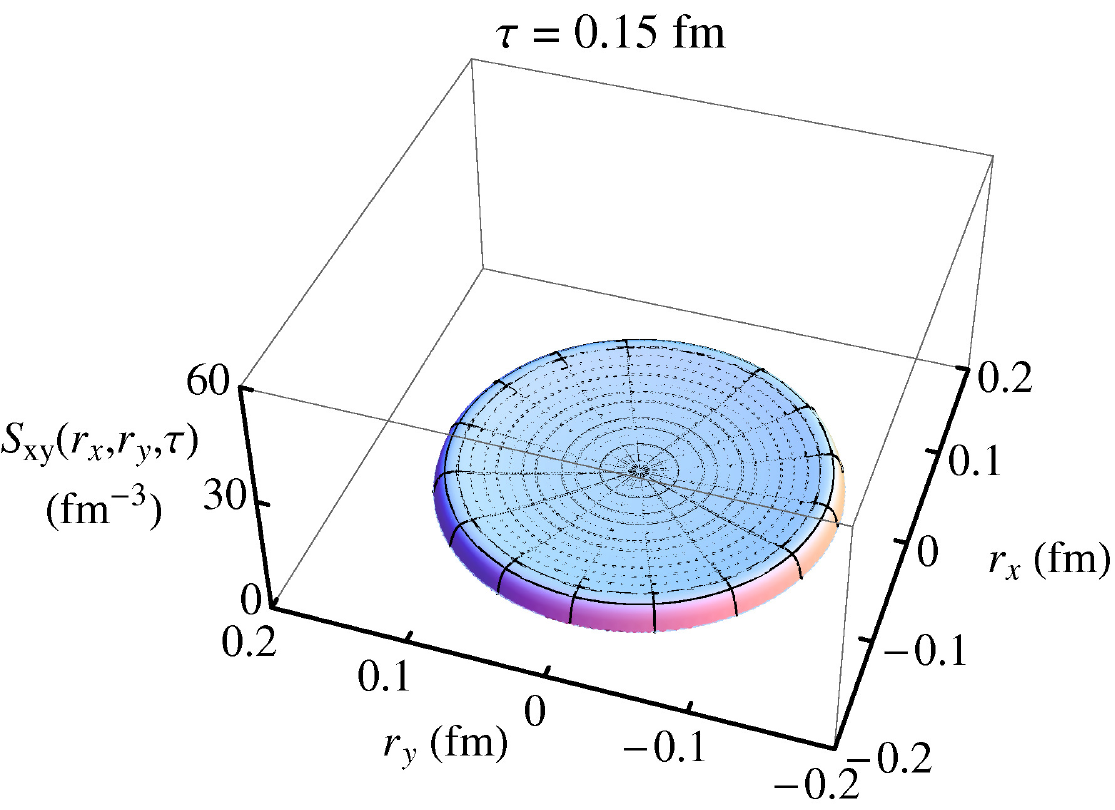} \includegraphics[width=5cm]{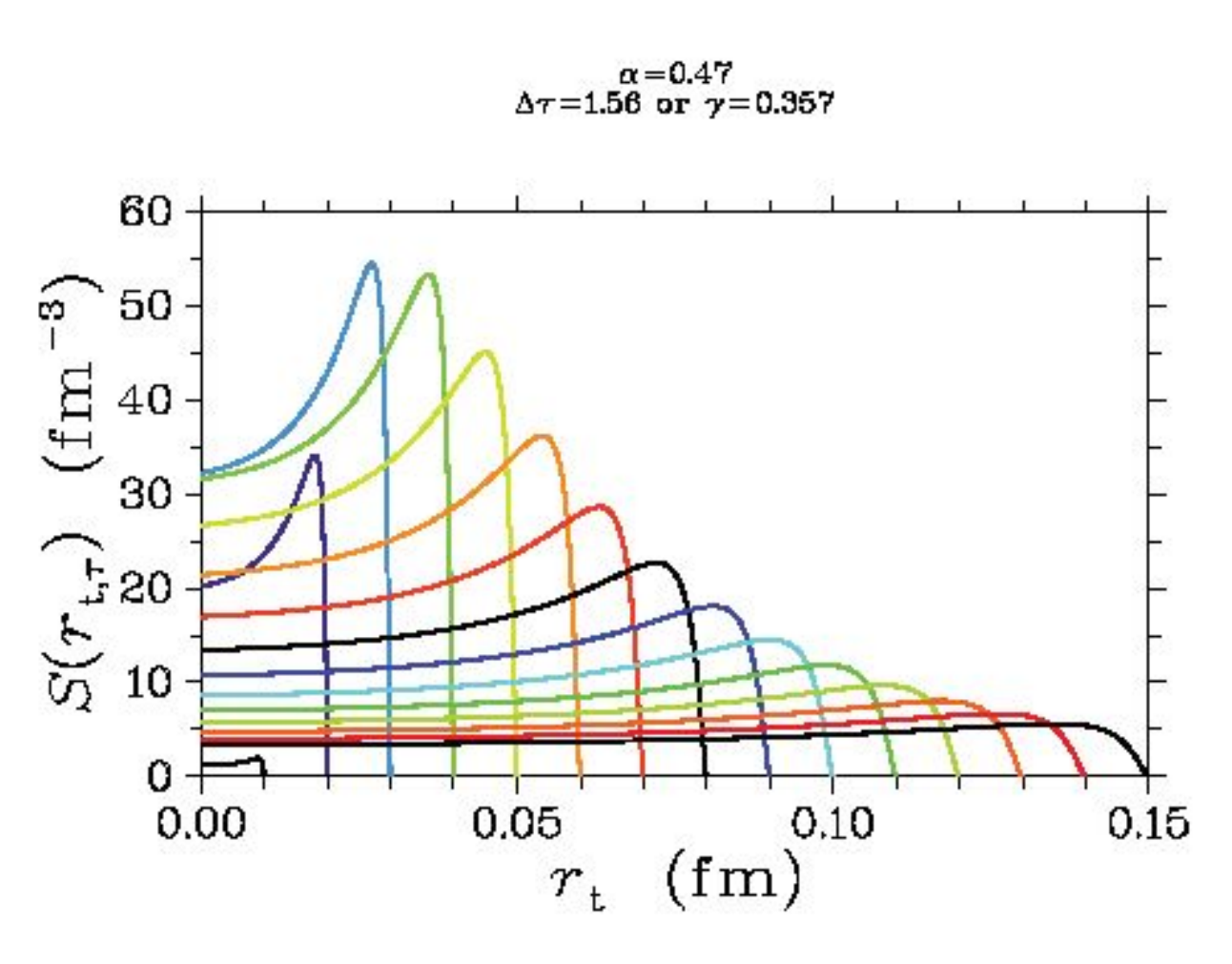} 
\caption{The transverse emission function normalized to unity, and its transverse profile for various 
proper times \cite{L3-Achard}. An animated gif file covering the first 0.15 fm=0.5$\times$10$^{-24}$ sec is available \cite{novak}.}
\end{figure}

The integral over the transverse distribution is shown
in Fig. 24. It exhibits a ''boomerang'' shape with a maximum at low $t$ and $z$,
but with tails reaching out to very large values of $t$ and $z$, a feature already
observed for hadron-hadron \cite{Agab1998,Ster2006} and heavy ion collisions \cite{Ster1999} (Fig. 25 a) and c))
in the framework of a hydrodynamical model \cite{Csor/9809011}.

The transverse part of the emission function is obtained by integrating over $z$ as well as azimuthal angle.
Fig. 26 shows
the transverse part of the emission function for various proper times. Particle
production starts immediately, increases rapidly and decreases slowly. In the
transverse direction, a ring-like structure is observed similar to the expanding,
ring-like wave created by a pebble in a pond. This ring like structure was also
observed in hadron-hadron collisions \cite{Agab1998} (Fig. 25 b)), where it was 
interpreted as due to the production of a fire-ring. Despite this similarity the 
physical process is different. Reflecting  a non-thermal nature of e$^+$e$^-$ annihilation, 
the proper-time distribution and space-time structure are reconstructed here 
without any reference to a temperature.

Interpolating and extrapolating Fig. 26, the proper-time dependence of the transverse 
expansion of the emission function can be best shown in a movie that ends in  
0.15 fm (0.5$\times$10$^{-24}$ sec), making it the shortest movie ever made of a process in 
nature \cite{novak}.

In conclusion, I find it absolutely amazing how the combination of experimental results on single particle spectra and two-particle correlations with some theoretical interpretation can allow us to construct a ''femtoscope'' and actually watch particle production at a scale below one fm taking place in less than 
10$^{-24}$ sec !

However, one basic puzzle remains: why in the world should pions thought to be produced {\em coherently} in a flux tube 
(at least in e$^+$e$^-$ and TeV pp collisions) be subject to {\em incoherent} Bose-Einstein correlations?
Have we fooled ourselves for the past half century or more? Perhaps! What Todorova-Nov\'a is trying 
to tell us in a series of papers \cite{todo14} based on the Lund Helix model \cite{anguha98} is just that. Bose-Einstein correlations may not be needed to explain the charge asymmetry of pion pair production, a helix shaped flux tube would not only generate transverse momenta and hadronic masses, but a sharp correlation peak for like charged pion pairs at low values of four-momentum difference $Q$. 
I think, it should be a fruitful challenge for younger ones among us to help sort that out in detail in the future. 

\vskip 2mm
\ul{​Acknowledgements.} 
I would like to thank first of all Andrzej, himself,
​for almost half a century of direct and indirect encouragement and
​guidance in an attempt to understand multihadron dyamics, and I would
like to thank Michal Prasza\l owicz and the organizers of this very special
Symposium for the honor of being invited to contribute.

\end{document}

The text...
\end{document}


\end{document}